\documentclass[12pt]{article}
\usepackage{amsmath}
\usepackage{graphicx}
\usepackage{enumerate}
\usepackage{natbib}
\usepackage{url}

\usepackage{authblk}

\usepackage {xr}
\makeatletter
\newcommand*{\addFiledependence}[1]{
  \typeout{(#1)}
  \@addtofilelist{#1}
  \IfFileExists{#1}{}{\typeout{No file #1.}}
}
\makeatother

\newcommand*{\myexternaldocument}[1]{
    \externaldocument{#1}
    \addFiledependence{#1.tex}
    \addFiledependence{#1.aux}
}

\myexternaldocument{app1013}
\numberwithin{equation}{section}

\usepackage{natbib,graphicx,setspace,lscape,longtable}
\usepackage{epsfig,epstopdf}
\usepackage{amsmath,amsthm,amssymb,color}

\newcommand{\BX}{\mathbf{X}}

\usepackage{hyperref} 
\usepackage{latexsym}
\usepackage{epsfig}
\usepackage{bm}
\usepackage{algorithm,algorithmic}
\usepackage{threeparttable}
\usepackage[small]{caption2}
\usepackage{threeparttable}
\usepackage{dcolumn}
\usepackage{multirow}
\usepackage{booktabs,epstopdf}

\newtheorem{theorem}{Theorem}[section]
\newtheorem{corollary}{Corollary}[section]

\newtheorem{proposition}{Proposition}[section]

\newtheorem{lemma}{Lemma}[section]

\newtheorem{assumption}{Assumption}

\begin{document}

\def\spacingset#1{\renewcommand{\baselinestretch}%
{#1}\small\normalsize} \spacingset{1}


{\small
\title{\bf Effective Positive Cauchy Combination Test}
  \author{Yanyan Ouyang\\
    Center for Applied Statistics and School of Statistics, Renmin University of China, \\
    School of Science, Chang’an University \\
    ~\\
    Xingwei Liu 
      \\
    Center for Applied Statistics and School of Statistics, Renmin University of China \\
    ~\\
    Lixing Zhu \\
    Center for Statistics and Data Science, Beijing Normal University\\ ~\\
    Wangli Xu \\
    Center for Applied Statistics and School of Statistics, Renmin University of China}
\maketitle
}







\begin{abstract}
In the field of multiple hypothesis testing, combining $p$-values represents a fundamental statistical method. 
The Cauchy combination test (CCT) \citep{cct} excels among numerous methods for combining $p$-values with powerful and computationally efficient performance.  
However, large $p$-values may diminish the significance of testing, even extremely small $p$-values exist. We propose a novel approach named the positive Cauchy combination test (PCCT) to surmount this flaw. Building on the relationship between the PCCT and CCT methods,  we obtain critical values by applying the Cauchy distribution to the PCCT statistic. We find, however, that the PCCT tends to be effective only when the significance level is substantially small or the test statistics are strongly correlated. Otherwise, it becomes challenging to control type I errors, a problem that also pertains to the CCT.  Thanks to the theories of stable distributions and the generalized central limit theorem, we have demonstrated critical values under weak dependence, which effectively controls type I errors for any given significance level. For more general scenarios, 
we correct the test statistic using the generalized mean 
method, which can control the size under any dependence structure and cannot be further optimized.
Our method exhibits excellent performance, as demonstrated through comprehensive simulation studies. We further validate the effectiveness of our proposed method by applying it to a genetic dataset.
\end{abstract}




\section{Introduction}

In practical scenarios, the necessity of multiple hypothesis testing becomes evident. For instance, in clinical medicine experiments, researchers assess the effectiveness of drugs by considering multiple factors, such as patient symptoms and physical indicators, each posing a separate hypothesis. Similarly, in genetics, researchers delve into microarray data to pinpoint genes that differ significantly, even though only a select few may show tangible differences. In such scenarios, conducting inferences through multiple hypothesis tests becomes essential. 
As a result, multiple hypothesis testing has become a vital methodology across a wide range of disciplines, serving as an indispensable tool for researchers and investigators to draw accurate conclusions.

Within the framework of multiple hypothesis testing, a series of null hypotheses are considered, denoted as $H_{0,1}, \ldots, H_{0, K}$, with corresponding alternative hypotheses labeled as $H_{1,1}, \ldots, H_{1,K}$.
These individual null hypotheses converge to form a collective global hypothesis expressed as $\boldsymbol{H}_0$ being the intersection of all $H_{0, i}$ for $i = 1, 2, \dots, K$, versus $\boldsymbol{H}_1$ being the union of all $H_{1, i}$ for $i = 1, \dots, K$.
To evaluate these hypotheses, statistics associated with the individual null hypotheses, $H_{0,i}$, versus the alternative hypothesis, $H_{1,i}$, are denoted as $X_i$ for $i=1,2, \ldots, K$, respectively. 
The associated $p$-values, labeled as $p_i = p\left(X_i\right)$, are employed to assess the validity of the hypotheses. The combination test statistic for the global null hypothesis $\boldsymbol{H}_0$ versus the alternative hypothesis $\boldsymbol{H}_1$ can be expressed as $T: = T(p_1,\ldots,p_K)$, where $T$ is a function defined over the interval $[0,1]^K$.

Since the introduction of Fisher's combination test statistic in \cite{Fisher}, there have been numerous studies and long-standing interests focused on multiple hypothesis testing. 
Early studies predominantly focused on deriving combination test statistics by summing transformed $p$-values. It was commonly assumed that these $p$-values were independent, facilitating the establishment of the statistical distribution under the null hypothesis. 
Among these methods, the seminal work by \cite{Fisher} proposed statistic  $-2\sum_{i=1}^K \mathrm{ln} p_i$ with a chi-squared distribution. \cite{Pearson} developed a statistic as $-\sum_{i=1}^K \mathrm{ln} (1-p_i)$ following a gamma distribution under the null, and \cite{Eugene} presented the approach of summing individual $p$-values as $\sum_{i=1}^K p_i$. 
These conventional methods work well with strong and non-sparse signals.

In the swiftly evolving field of big data, many domains are challenged with combining $p$-values with weak and sparse signals. In such contexts, the aforementioned methods may suffer substantial power loss. A variety of methods have been developed to address this issue. To name a few, early work include $1 - (1-p_{(1)})^K$ proposed by \cite{1931methods} and $K p_{(1)}$ for the Bonferroni method detailed in \cite{1961Bonferroni}, utilizing order statistics for multiple testing.
Later,  Berk-Jones test presented in \cite{Berk1979GoodnessoffitTS} has shown exemplary performance in terms of Bahadur efficiency. Another notable innovation is the higher criticism statistic by \cite{Donoho2004HigherCF}, which effectively detects weak and sparse signals and represents a cornerstone in the field.
To our knowledge, however, these methods primarily addressed independent $p$-values, thus providing no analytic formulas for correlated $p$-values. Techniques such as permutation are computationally burdensome or even at times infeasible, especially when the $p$-value of a combination test is extremely small. 
Methods like \cite{1986simes} are valid only under moderate dependence assumptions, which is often violated in practical scenarios, as indicated by \cite{Efron2010LargescaleI} on page 51. 
This underscores the need for more adaptable and robust methods capable of handling complex dependencies in data.

Recently, to address these challenging situation from genome-wide association studies, where genetic data may contain million of genetic variants and computational efficiency is a necessity,  \cite{cct} proposed the Cauchy combination test (CCT) to combine $p$-values under an unspecified dependence structure. 
The CCT statistic is denoted as 
\begin{equation}\label{eq01}
    T_{\rm {CCT}}: = \sum_{i=1}^{K} \frac{1}{K} \tan\{ \left(0.5-p_i\right) \pi\}.
\end{equation}
The study presented in \cite{cct} revealed that the tail probability of $T_{\rm {CCT}}$ can be approximated by a standard Cauchy distribution under the null hypothesis, assuming bivariate normality for a fixed $K$ and additional two assumptions as $K$ diverges. Due to its simplicity, user-friendliness, and powerful effectiveness, the Cauchy combination test has garnered significant recognition and citations since its inception, as evidenced by references such as \cite{2023Zhao, 2023Yu, 2023Long}. Distinct from alternative methodologies, the approach in \cite{cct} utilizes  $p$-value transformation method employing $\tan\left\{(0.5-p_i)\pi\right\}$ for individual $p_i$. This technique notably accentuates the significance of smaller $p$-values and potentially improves the detection of weak and sparse signals.

However, the function $\tan\left\{(0.5-p_i)\pi\right\}$ yields opposite values at $p_i$ and $1-p_i$. A consideration arises when a $p$-value is very close to one, where the contribution in the $T_{\rm {CCT}}$ in \eqref{eq01} is close to negative infinity. 
Even though the $p_i$ is extremely small, it is still hard to reject the null hypothesis with a large $p$-value such as $1-p_i$. This large negative penalty may diminish the significance of individual $p$-value, leading to a substantial power loss.
To elucidate this issue clearly, consider the case with $K=2$, where the individual $p$-values are $p_1=0.001$ and $p_2=0.999$. Even employing the Bonferroni method, which is renowned for a conservative approach, we can reject the null hypothesis at any significant levels greater than 0.002. However, due to the large penalty of $\tan\left\{(0.5-p_2)\pi\right\}$, the combined test statistic $T_{\rm{CCT}}$ equals to 0, lacking of evidence to reject the null.

The above scenario seems hard to happen since valid $p$-values under the null distribute on $U(0,1)$, uniform distribution on $[0,1]$, while under the alternative, they are stochastically smaller than $U(0,1)$, which implies that the $p$-values approach to 0 with higher probability. However, large $K$ may lead to $p$-values close to one. 
It also may occur if the alternative hypothesis does not include the true scenario.
This issue is also considered in \cite{Fang2021HeavyTailedDF}, which proposed a truncated Cauchy test to alleviate the problem. But its main flaw is the subjective selection of the truncated value, lacking theoretical supports. 

To address the issue of the transformation of $p$-values with the large negative penalty in CCT, we propose a novel combined test statistic, referred to as the positive Cauchy combination test (PCCT) and denoted as 
\begin{eqnarray} \label{eq02}
T_{\rm {PCCT}}: = \sum_{i=1}^{K} \frac{1}{K} \tan \left\{\left(0.5-p_i/2\right) \pi\right\}.
\end{eqnarray} 
A key advantage of this approach is that all the summation terms in $T_{\rm {PCCT}}$ are guaranteed to be non-negative, in contrast to the statistic $T_{\rm {CCT}}$. 
Compared to the CCT method, the transformation of $p$-values to non-negative values avoids the potential mutual cancellation of different $p$-value transformations.
Therefore, $T_{\rm {PCCT}}$ is significantly less affected by extremely large $p$-values, providing a potential for power to rival $T_{\rm {CCT}}$, especially in scenarios characterized by sparse and weak signals.

An intriguing perspective can shed light on a deeper understanding of one of the motivations behind \eqref{eq02} and provide insights from another viewpoint. 
Notice that a connection can be built between transformation $\tan\{(0.5-p_i)\pi \}$ in \eqref{eq01} and a one-sided test. Specifically, suppose we aim to test $H_0:\mu=0$ versus $H_1:\mu>0$, using an observation $x$ drawn from a population following $\rm{Cauchy}(\mu,1)$. The $p$-value is $p=\int_x^{\infty}1/\{(1+y^2)\pi\}dy$, and $x$ can be solved as $x=\tan\{(0.5-p)\pi\}$. It may be the rational that motivates the CCT.
When testing the two-sided hypothesis with the alternative  $H_1:\mu\neq0$, the two-sided $p$-value is given by $p=2\int_{|x|}^{\infty}1/\{(1+y^2)\pi\}dy$. This results in $|x|=\tan\{(0.5-p/2)\pi\}$, which corresponds exactly the proposed test statistic $T_{\rm {PCCT}}$ in \eqref{eq02}. 
In the other words, the original CCT can be viewed as a kind of one-sided test, while the proposed PCCT is designed for the two-sided test. As the result, the PCCT inherently enjoys robustness with large $p$-values close to one.

To establish the critical value for the statistic $T_{\rm {PCCT}}$, a relationship between the distribution of statistics $T_{\rm {CCT}}$ and $T_{\rm {PCCT}}$ under the null hypothesis is established. Note that the Cauchy distribution's tail probabilities are better approximated for $T_{\rm {CCT}}$ at a smaller significance levels, or with weak or strong dependence. A similar phenomenon, the accuracy decreases as the significance level increases without strong dependence, exists for $T_{\rm {PCCT}}$, which will be further elaborated upon in detail in Section \ref{sec:corrected}. 
Therefore, the approximation may not work well, especially at larger significance levels. 
To address this issue, we derive the limit distributions under weak dependence with the aid of stable distribution and generalized central limit theorem. Further, drawing inspiration from \cite{vovk2020} constructing generalized mean for $p$-values, we propose a corrected critical region for the statistic $T_{\rm {PCCT}}$ which effectively controls type I errors without any additional assumptions. 
Numerical analysis results indicate that our proposed method exhibits significant superiority over some existing methods.

In words, our work has distinguished contributions in three aspects. First, to conquer the negative penalty that CCT may encounter, we propose a novel method named PCCT, which meanwhile remains the advantages of the the CCT. 
Second, when deriving the critical values of the statistic, we discuss two conditions, including bivariate normality and weak dependence, and finally all the conditions are removed. The theoretical results are established under each conditions. Further, the methodology is adopted to derive corresponding critical values of the CCT and the harmonic mean method. 
Third, the asymptotic equivalence of these three combination methods is investigated in deep, which is demonstrated both theoretically and numerically. 

The remainder of the paper is structured as follows. In Section \ref{sec2}, we discuss the connection between the existing CCT and the proposed PCCT, and its non-asymptotic distribution is derived under mild assumptions. 
In Section \ref{sec:corrected}, we derive the limit distributions of the PCCT and the CCT under weak dependence assumption, and further obtain corrected and conservative critical regions of the PCCT when all the assumptions are removed.
The asymptotic properties and the efficiency of the proposed approach are discussed in Section \ref{sec:property}. 
Section \ref{sec:simulation} consists of a part of simulation study to assessing the validity and power of the proposed method compared to other tests. Conclusions are drawn in Section \ref{sec:conclusion}, and theorem proofs along with additional numerical analysis are provided in the Supplemental Materials.

\section{ 
The Connection between CCT and PCCT Statistics
}\label{sec2}


Here, we first review the Cauchy combination test in (\ref{eq01}).  
For fixed $K$, \cite{cct} assume that the $p$-values are obtained from z-scores, that is, the CCT statistic in \eqref{eq01} can be characterized by
\begin{eqnarray}
\label{eq03} 
T_{\rm{CCT}} = \sum_{i=1}^{K} \frac{1}{K}\tan \left[\left\{2\Phi(| X_i| )-3/2\right\} \pi\right].
\end{eqnarray}
where $X_i$ is a z-score corresponding to the individual $p$-value $p_i$, and $\Phi(\cdot)$ denotes the cumulative distribution function of a standard normal variable. 
We assume each $X_i$ has variance 1. The null $\boldsymbol{H}_0$ can be formulated as $H_{0,i}:E(X_i)=0$ for $i=1,2,\ldots,K$, while the alternative is $H_{1,i}:E(X_i)\neq0$ for some $i$. Assuming that bivariate normality condition about the $X_i$'s, they demonstrated the tail probability of the $T_{\rm{CCT}}$ could be well approximated by a standard Cauchy distribution under the null. 
It is a shocking result that the tail probability of the $T_{\rm{CCT}}$ remains almost unchanged as the dependence of $X_i$ varies. 
Noting that $\tan \left\{\left(0.5-p_i\right) \pi\right\}$ following the standard Cauchy distribution, which is a classical heavy-tailed distribution, this surprising result can be explained intuitively as follows. 
The bivariate normality essentially guarantees that the tail distributions of each pair of $X_i$ and $X_j$ are asymptotically independent. Therefore, the dependence structure of each pair of $p_i$ and $p_j$ has limited influence on $T_{\rm{CCT}}$. 
In other words, the interplay between the heavy-tailed transformation distribution and tail independence limits the impact of the dependence structure, leading to the robustness of $T_{\rm{CCT}}$ under arbitrary dependence structure.

We present that the proposed PCCT outperforms the CCT in two main aspects. On the one hand, aforementioned discussion indicates that a crucial advantage of the CCT, the heaviness of the tail distribution, is also exhibited by the proposed PCCT. 
It is sufficient to note that $\tan \left\{\left(0.5-p_i/2\right) \pi\right\}$ shares the same distribution with the absolute value of a standard Cauchy variable under the null, processing a heavier tail than the standard Cauchy distribution. Therefore, it is foreseeable that the PCCT is also robust for the dependence structure. 
Moreover, another advantage of the heavy-tailed distribution is its ability to amplify small $p$-values, whereas pay little attention to the large ones, thus detecting sparse signals efficiently. 
On the other hand, it is noted that $p$-values greater than $0.5$ negatively contribute to the CCT. Extremely large $p$-values lead to a cancellation effect between the positives and negatives, reducing the power of the test.
A slight modification can work this issue out. The PCCT in \eqref{eq02} ensures that each $p$-value in the sum remains positive after transformation, diminishing the negative penalty. In a word, this small modification remains the advantages of the CCT, while works the negative penalty issue out. 

Next, we formally formulate the bivariate normality assumption as well as the non-asymptotic approximation to the null distribution of $T_{\rm{PCCT}}$. 
Assume that the statistic can be characterized by z-scores, i.e.
\begin{eqnarray}
    T_{\rm{PCCT}} = \sum_{i=1}^{K} \frac{1}{K}\tan \left[\left\{\Phi(|X_i| )-1/2\right\} \pi \right].
\end{eqnarray}

\begin{assumption}\label{ass1}
    For any $1\leq i<j\leq d$, $(X_i,X_j)^\top$ follows a bivariate normal distribution.
\end{assumption}

Assumption \ref{ass1} is also assumed in \cite{cct}, \cite{Efron2007CorrelationAL}, which is utilized for guaranteeing that each pair are asymptotically tailed independent; see \cite{Fang2021HeavyTailedDF} for details. As pointed out in \cite{2023Long}, this assumption is too stringent for practical applications, and it can be weaken as $X_i$ and $X_j$ have an arbitrary bivariate distribution instead. However, this extension is far from our main line, thus we keep Assumption \ref{ass1} throughout this section. The following theorem provides a non-asymptotic approximation to the null distribution of $T_{\rm{PCCT}}$. 

\begin{theorem}\label{theorem:vsd}
Suppose Assumption \ref{ass1} holds and $E(X_i) = 0, 1\le i\le K$. 
Then for fixed $K$, we have
\begin{eqnarray}
\label{eq:pcct}
\lim_{t\to+\infty}\frac{{\rm pr}(T_{{\rm PCCT}} > t)}{2{\rm pr}(W_{\mathcal{C}}> t)} = 1, 
\end{eqnarray}
where $W_{\mathcal{C}}$ represents a standard Cauchy random variable. 
\end{theorem}

Theorem \ref{theorem:vsd} indicates that $T_{\rm{PCCT}}$ shares the same tail distribution with the absolute value of a standard Cauchy variable. 
It is interesting to note that $T_{\rm{PCCT}}$ is an average of dependent absolute values of standard Cauchy variables under the null, and the average keeps the similarity with the absolute value of a standard Cauchy consistently no matter how the dependence structure varies. 
Roughly speaking, the heaviness of the tail of Cauchy distribution leads to limited impacts of the correlated structure. Further, the critical value of $T_{\rm{PCCT}}$ can be obtained by the standard Cauchy distribution. Theorem \ref{theorem:vsd} suggests that an $\alpha$-level test can be rejected if and only if $T_{\rm{PCCT}}>t_{\alpha/2}$, where the $t_{\alpha/2}$ is the upper $\alpha/2$-quantile of the standard Cauchy distribution. The $p$-value of the test can be simply approximated by
\begin{eqnarray}\label{eq04}
    p=1-2\arctan(T_{\rm PCCT})/\pi,
\end{eqnarray}
here $T_{\rm{PCCT}}$ represents a realization of the PCCT statistic. 

Theorem \ref{theorem:vsd} in \cite{cct} also presented similar results for $T_{\rm{CCT}}$. It is worth pointing out that, based on these two theorems, the power of $T_{\rm{PCCT}}$ appears to consistently precede that of $T_{\rm{CCT}}$.
We are to state this matter formally in Theorem \ref{theo:power} (i), and reveal the underlying essence following the theorem. 

We close this section by pointing out that Theorem \ref{theorem:vsd} restricts fixed combination number $K$, which can be relaxed to $K$ diverging under some mild conditions and order limitation between $K$ and $t$. But it is far from our storyline, thus omitted here. 

\section{The Corrected Critical Region}\label{sec:corrected}
Both Theorem \ref{theorem:vsd} in this paper and Theorem 1 in \cite{cct}
imply that the accuracy of approximating the tail probability of the statistics improves as the $t$ increases, or equivalently, as the $\alpha$ decreases.
However, for any given significance level, these approximations may not be precise enough, especially for the CCT with moderate dependence or the PCCT without strong dependence, leading to inflation of the sizes. To elucidate this point clearly, a simple numerical analysis is presented below. 

We sample $\boldsymbol{X}=(X_1, \dots, X_K)^\top$ from a multivariate normal distribution $N(\boldsymbol \mu,\boldsymbol \Sigma)$, where $\boldsymbol\Sigma= (1-\rho) \boldsymbol I_K+\rho \boldsymbol J_K$ and $K=10^3$. Here, $\boldsymbol I_K$ and $\boldsymbol J_K$ denote the $K$-dimensional identity matrix and a $K$-dimensional matrix with all elements are $1$, respectively, and $\rho=0, 0.01, $ $0.2, 0.3, 0.5, 0.7, 0.99$. Hypothesis tests of interest can be formulated as null hypothesis $H_{0, i}: \mu_i=0$, where $\mu_i$ represents the $i$-th element of $\boldsymbol{\mu}$. The $p_i$'s can be calculated by $2\{1-\Phi(|X_i|)\}$. Given significance levels $\alpha=0.05, 0.01, 0.001$, the empirical sizes of the global null hypothesis $\boldsymbol{H}_0$ are calculated through $10^6$ repetitions. The results are summarized in Table \ref{table1}.
\begin{table}[ht]
\caption{Empirical sizes for the CCT and the PCCT with varying $\alpha$ and $\rho$}
\centering
\begin{tabular}{ccccccccccc}
  \toprule
  & &\multicolumn{8}{c}{$\rho$} \\
  \cmidrule(lr){3-10}
$\alpha$ & method & 0 & 0.01 & 0.1 & 0.2 & 0.3 & 0.5 & 0.7 & 0.99  \\ 
  \midrule
0.05 & CCT & 0.0492 & 0.0567 & 0.1009 & 0.1110 & 0.1042  & 0.0838  & 0.0667 &  0.0515 \\ 
 & $\rm{PCCT}$ & 0.0894 & 0.0948 & 0.1267 & 0.1229 & 0.1107 & 0.0859 & 0.0673 & 0.0511 \\
 \midrule
0.01 & CCT & 0.0099 & 0.0102 & 0.0134 & 0.0171 & 0.0190  & 0.0171 & 0.0135   & 0.0102 \\
& $\rm{PCCT}$ & 0.0113 & 0.0120 & 0.0145 & 0.0174 & 0.0185 & 0.0169 & 0.0140 & 0.0101 \\  
 \midrule
0.001 & CCT & 0.0011 & 0.0010 & 0.0012 & 0.0014 & 0.0014  & 0.0015  & 0.0013   & 0.0011 \\ 
& $\rm{PCCT}$ & 0.0010 & 0.0011 & 0.0012 & 0.0013 & 0.0014 & 0.0016 & 0.0014 & 0.0009 \\ 
   \bottomrule
\end{tabular}
\label{table1}
\end{table}

The CCT method can rather effectively control the type I error when the statistics are independent ($\rho=0.0$) or highly correlated ($\rho=0.99$), which aligns with similar results in \cite{Fang2021HeavyTailedDF}. 
In the scenarios where the $p$-values of individual statistics are either independent or perfectly correlated, corresponding to $\rho=0$ and $\rho=1$, respectively, the exact distribution of the $T_{\rm{CCT}}$ follows a standard Cauchy distribution. Meanwhile, the proposed $T_{\rm{PCCT}}$ precisely follows the distribution function $F_{\mathcal{P}}(u)=2\arctan(u)/\pi$ precisely when $\rho = 1$, thus the size can be well controlled with highly correlated statistics. 
However, both the PCCT under weak dependence and the CCT with moderate correlations, such as $0.1 \leq \rho \leq 0.7$, suffer explosions of the empirical sizes, especially at larger significance levels. 
In other words, without enough small significance levels, both the CCT with moderate dependence and the PCCT without strong dependence may encounter the inflation of the type I error rate.


Based on the above analysis, before delving into more general scenarios, it is evident that obtaining a better approximation of the test statistic is crucial under the assumption of weak dependence, especially for the proposed PCCT. 
This result is primarily based on the theories of stable distributions and the generalized central limit theorem, as referenced in \cite{1994Stable, 1999StableDistributions}.

\begin{assumption}\label{ass2}
Assume that the sequence $\{X_i\}$ satisfying 
\begin{align}
    \sup_{A\in \sigma(X_1,\ldots,X_{k}), B\in \sigma(X_{k+h},\ldots,X_{K})} |{\rm pr}(A\cap B) - {\rm pr}(A){\rm pr}(B)|:=\alpha_h
   \to 0,\quad h\to\infty.
\end{align}
where $h$ satisfying $ h/K \to 0$ and $K\alpha_{h} \to 0$.
\end{assumption}

Assumption \ref{ass2} describes a kind of weak dependence, which promises the correlation of $X_i$ and $X_j$ tends to zero as the distance $|i-j|$ increases. It is a short-term dependence which holds in many cases such as the effects of genetic variants. 
In familiar terms, Assumption \ref{ass2} suggests the sequence $\{X_i,i=1,\ldots, K\}$ is strongly mixing with the rate function $\alpha_h$ decaying fast. 
The strongly mixing assumption has been widely utilized to investigate hypothesis testing problems (\cite{gregory2015two,Hu2018Diagonal}).
The following theorem presents the limit distribution of $T_{\textrm{CCT}}$ and $T_{\textrm{PCCT}}$ under Assumptions \ref{ass1}-\ref{ass2}. 

\begin{theorem}\label{th:weak} 
Suppose Assumptions \ref{ass1}-\ref{ass2} hold and $E(X_i) = 0, 1\le i\le K$. Then we have
\begin{align}
T_{\rm CCT}  \stackrel{d}{\to} W_{\mathcal{C}}\qquad {\rm as}\ K\to \infty,
\end{align}
where $W_{\mathcal{C}}$ follows a standard Cauchy distribution, and
\begin{align}\label{eq:ind}
T_{\rm PCCT}-\Delta_{\mathcal{P}}\stackrel{d}{\to} S_0\qquad {\rm as}\ K\to \infty.
\end{align}
Here, the characteristic of $S_0$ is $\chi(t) = \exp\{-|t|-2i{\rm sign}(t)\log|t|/\pi\}$, and
\begin{align}
\Delta_{\mathcal{P}} = K\mathbb{E}(W_{\mathcal{P}}/K) = K\int_{0}^{\infty}\sin(x/K)\frac{2}{\pi (x^2+1)}dx, 
\end{align}
where the distribution function of $W_{\mathcal{P}}$ is $F_{\mathcal{P}}(x) = 2\arctan(x)/\pi$. 
\end{theorem}

By Theorem \ref{th:weak}, the $\alpha$-level null is rejected if and only if $T_{\textrm{PCCT}} \geq p_S(1-\alpha)+\Delta_{\mathcal{P}}$ for the PCCT, and $T_{\textrm{CCT}} \geq p_{C}(1-\alpha)$ for the CCT,  where $p_S(1-\alpha)$ and $p_C(1-\alpha)$ denote the $(1-\alpha)$-quantile of $S_0$ and $W_{\mathcal C}$, respectively. 
However, Theorem \ref{theorem:vsd} constrains the $K$ to be fixed,
or $K^2/t\to 0$ at most, 
while Theorem \ref{th:weak} holds for $K$ large enough. Consider a fixed dataset with $K=40$ at level $\alpha=0.001$, corresponding $t=318.30$. Whether the practitioners should hold that $K^2/t\to0$ or not leaves them in a bind. 
Such assumptions are absolutely unverifiable since we typically have only one dataset at hand. Similar discussions are also presented in \cite{Kim2020DimensionagnosticIU}, which considered the goal named dimension-agnostic inference. 
The restriction of $K$ in the both theorems motivates us to develop results without such assumptions. 

Therefore, for the two purposes, controlling the size precisely under arbitrary dependence and removing the constraints on $K$, from above simulations and discussions, we rearrange the critical region through the application of the inverse transformation function of the $p$-value function. This technique is valid under arbitrary dependence structures, and also makes no assumptions about $K$. From here to the end of this section, we pay all attention to the proposed PCCT. 

We  first introduce the concept of the generalized mean of $p$-values. As proposed by \cite{1930Kolmogorov}, a generalized mean of $p$-values can be written as
\begin{eqnarray}\label{eq31}
M_{\phi}(p_1,\ldots,p_K) = \psi\left \{\frac 1 K \sum_{i=1}^K \phi(p_i) \right\}.
\end{eqnarray}
Here, $\phi:[0,1]\to [-\infty, \infty]$  represents a continuous strictly monotonic function, while $\psi:\phi([0,1]) \to [0,1]$ serves as its inverse. 
A specific class of the generalized mean of $p$-values discussed by \cite{vovk2020} is $\phi_r(p) = p^r$, where $r\in \mathbb{R}\cup \{-\infty,\infty\}$. 
Type I error is controlled by multiplying the critical value with different correction coefficients in various cases, such as the harmonic mean and the arithmetic mean, corresponding to $r = -1, 1$ respectively. 
Specifically,  \cite{vovk2020} consider 
a constant $C$ satisfying
${\rm pr}\left(M_{\phi}(p_1,\ldots,p_K) \leq \alpha/C\right) \leq \alpha$, rejecting the null under $\alpha$-level when $M_{\phi}(p_1,\ldots,p_K) \leq \alpha/C$. 
More generally, we consider a continuous increasing function $g:(0, 1) \to[0,1]$ satisfying
\begin{align}\label{eq:g}
{\rm pr}\left(M_{\phi}(p_1,\ldots,p_K) \leq g(\alpha)\right) \leq \alpha.
\end{align}
Then the null hypothesis can be rejected if $M_{\phi}(p_1,\ldots,p_K) \leq g(\alpha)$.

Before deriving the  critical value based on aforementioned concept, we introduce some notations. 
Let $\mathcal{U}$ be the set of all random variables distributed uniformly over the interval $[0,1]$, then $p_1,\ldots,p_K \in \mathcal{U}$ under the null. Let the left $\alpha$-quantile as $q_\alpha(X):=\sup \{x \in \mathbb{R}: {\rm pr}(X \leq x)<\alpha\}$. For $\alpha \in(0,1)$,
the threshold $a_{\phi}$ of the generalized mean $M_{\phi}$ is defined as
\begin{align}
&a_{\phi}(\alpha):=\inf \left\{q_\alpha\left\{M_{\phi}(p_1,\ldots,p_K)\right\} \mid p_1,\ldots,p_K \in \mathcal{U}\right\}.
\end{align}
Exactly speaking, this threshold is valid for arbitrary dependence, abbreviated as VAD, so we call $a_{\phi}(\alpha)$ the VAD threshold of the generalized mean $M_{\phi}$. An intuitive lemma shows the relationship between $a_{\phi}(\alpha)$ and $g(\alpha)$ in \eqref{eq:g}. 

\begin{lemma}\label{lemma01}
For a generalized mean $M_{\phi}: [0, 1]^K \to[0,1]$ and continuous increasing function $g:(0, 1) \to[0,1]$, for any  $\alpha\in(0,1)$, \\
(a) $\sup \left\{{\rm pr} \left( M_{\phi}(p_1,\ldots,p_K)\leq g(\alpha) \right) \mid p_1,\ldots,p_K\in \mathcal{U}\right\} \leq \alpha$ if and only if $g(\alpha)\leq a_{\phi}(\alpha)$; \\
(b) $\sup \left\{{\rm pr} \left( M_{\phi}(p_1,\ldots,p_K)\leq g(\alpha) \right) \mid p_1,\ldots,p_K\in \mathcal{U}\right\} = \alpha$ if and only if $g(\alpha)=a_{\phi}(\alpha)$.
\end{lemma}

The conclusion (b)  in Lemma \ref{lemma01} implies that $g(\alpha) = a_{\phi}(\alpha)$ reaches the sharp bound, which can control the size under arbitrary dependence structure and can not be improved anymore. 
The VAD threshold of generalized mean has been discussed in recent studies with the results of robust risk aggregation, please refer to \cite{vovk2020, Chen2020TradeoffBV} for more details.

With this powerful technique, we now return to our focus. The function $\phi(p) := \phi_{\rm ptan}(p) = \tan\{(0.5 - p/2)\pi\}$ is utilized for the generalized mean in \eqref{eq31}, and the inverse function $\psi_{\rm ptan}(\cdot)$ can be derived as $\psi_{\rm ptan}(u) = 1 - 2{\rm arctan}(u)/\pi$. 
The generalized mean based on the function $\phi_{\rm ptan}(p)$ can be expressed as 
\begin{eqnarray}\label{eq32}
M_{\rm{ptan}}:=M_{\rm{ptan}} (p_1,\ldots,p_K) = 1-\frac{2}{\pi}{\arctan}\left[\sum_{i=1}^K\frac{1}{K}\tan\left\{(0.5-p_i/2)\pi\right\}\right].
\end{eqnarray}
Note that $M_{\rm{ptan}} = \psi_{\rm{ptan}}(T_{\rm{PCCT}})$, the proposed method can be viewed as a transformation of the generalized mean. Our focus is to calculate the VAD threshold $a_{\mathcal{P}}(\alpha)$ of $M_{\rm{ptan}}$ to obtain the sharp bound of $g(\alpha)$, thus controlling the size while remaining the power as possible. 
With the techniques in the field of robust risk aggregation in \cite{2013wangVaR}, we have the following proposition.

\begin{proposition}\label{prop:VAD} 
Let $a_{\mathcal{P}}(\alpha)$ be the VAD threshold of $M_{{\rm ptan}}$ for significance level $\alpha \in (0,1)$,  we have
\begin{align}
a_{\mathcal{P}}(\alpha) = \psi_{\rm{ptan}}(H_\alpha(x_K)/K),
\end{align}
where $H_\alpha(x) = (K-1)F_{\mathcal{P}}^{-1}(1-\alpha+(K-1)x) + F_{\mathcal{P}}^{-1}(1-x)$ with $x\in(0,\alpha/K)$, and $x_K$ is the unique solution to the equation
$K\int_x^{\alpha/K}H_\alpha(t){\rm d}t = (\alpha-Kx)H_\alpha(x)$, $F_{\mathcal{P}}(x) = 2\arctan(x)/\pi$.
\end{proposition}

Proposition \ref{prop:VAD} states that we can correct the critical region to control the type I error of the proposed method under arbitrary dependence structure. The null hypothesis is rejected if and only if $M_{\rm{ptan}} \leq a_{\mathcal{P}}(\alpha)$ or equivalently $T_{\rm{PCCT}} \geq \psi_{\rm{ptan}}(a_{\mathcal{P}}(\alpha))$.
Although our main proposal is to lift the restriction of $t\to+\infty$ in Theorem \ref{theorem:vsd}, thus controlling the size better, it is worth pointing out that Proposition \ref{prop:VAD} removes all assumptions such as bivariate normality in Assumption \ref{ass1}, $K$ be fixed or limited by $t$ as byproducts. 
That is to say, this corrected region is valid without any additional assumptions, and it is unable to improve anymore under such an extremely loose condition. 

We further proved that $\alpha/\{\log (K)a_{\mathcal{P}}(\alpha)\}\to 1$ as $K\to\infty$ in the proof of Proposition \ref{prop:VAD}. 
Then $a_{\mathcal{P}}$ can be approximated by $\alpha/\log (K)$. However, the rate convergence is very slow, and $\alpha/\{\log (K)a_{\mathcal{P}}(\alpha)\}> 1$ for moderate values of $K$. 
It is suggested to use the conservative thresholds in practical scenarios. Table \ref{table:af} in the Supplementary Materials reports numerical values of $\alpha/\{\log (K) a_{\mathcal{P}}(\alpha)\}$. For instance, for $K\geq100$, one may use $\alpha/(1.62\log (K))$ as the approximation of $a_{\mathcal{P}}$.

With several thresholds in hands, a natural issue comes to which one to choose in practical scenarios. We recommend the thresholds derived by Theorem \ref{th:weak} as a preferable choice without evidence against weak dependence, otherwise the VAD threshold is more safe but conservative. The choice of these thresholds is, in fact, a trade-off between validity and efficiency, or in other words, a trade-off between assumptions and the rejection region. The stronger the assumptions, the larger the rejection region. The VAD thresholds pay a significant price for its validity under such an extremely loose condition. Weak dependence, on the other hand, strikes a better balance between the two, preserving the power of tests under relatively mild conditions.

\section{Properties of the Test Statistics}\label{sec:property}
In this section, we are dedicated to the properties of the proposed PCCT. We first connect the PCCT with another combining methods, the CCT in \cite{cct} and the harmonic mean $p$-value (HMP) in \cite{2019HMP}, which both can combine $p$-values under an unspecified dependence structure, and further verify their asymptotic equivalence. 

As pointed out in \cite{Goeman2019TheHM} and \cite{Chen2020TradeoffBV}, both of those two methods are too aggressive. Therefore, one of our interests is comparing their VAD thresholds, where the CCT and the HMP thresholds are available in \cite{Chen2020TradeoffBV} and \cite{vovk2020}, respectively. To this end, these methods are characterized by generalized means first. Formally speaking, let $\phi(p) := \phi_{-1}(p) = 1/p$ in \eqref{eq31} correspond to the HMP, while $\phi(p) := \phi_{\rm{tan}} = \tan\{(1-p)\pi\}$ for the CCT method, then they are characterized by generalized means. Theorem \ref{theo:r=-1} reveals that the connection among the harmonic mean $M_{-1}$, the CCT method $M_{\tan}$ and the proposed method $M_{\rm{ptan}}$. Before proceeding, we denote $A_x \sim B_x$ as $x\to x_0$ to represent that $A_x/B_x \to 1$ as $x\to x_0$. 

\begin{theorem}\label{theo:r=-1} 
If $\max_{i=1,\ldots,K} p_i \le C$ for some fixed $C\in(0,1)$. For a fixed $K\in \mathbb{N}$ and any realizations of $p_i$'s, $M_{-1}$, $M_{\tan}$, and $M_{\rm{ptan}}$ are asymptotically equivalent as the minimum order statistic of $p$-values, $p_{(1)}$, tends to $0^+$, i.e. 
\begin{align}
M_{\rm{ptan}}\sim M_{-1}\sim M_{\tan}. 
\end{align}
\end{theorem}

Theorem \ref{theo:r=-1} asserts that $M_{\rm ptan}$ converges towards $M_{-1}$ and $M_{\tan}$ as $p_{(1)}\to 0^+$. It is surprising that even though the forms of these three generalized means are extremely different, they are asymptotically equivalent 
when an exceptionally small $p$-value is present. 
The similar result 
has also been demonstrated in \cite{Chen2020TradeoffBV} and \cite{Fang2021HeavyTailedDF}. 
As their generalized means are quite similar, it is natural to consider whether their critical regions also share the similarity. To verify our conjecture comprehensively, two kinds of thresholds are investigated. Similar to the VAD threshold, the threshold valid for weak dependence in Assumption \ref{ass2}, abbreviated as VWD, is defined as $b_{\phi}(\alpha)$ satisfying ${\rm pr}\{M_{\phi} \le b_{\phi}(\alpha)\} = \alpha$.
It represents the most aggressive valid bound for the weak dependence, and the approximate VWD threshold for PCCT and CCT can be directly calculated by Theorem \ref{th:weak}. 
Furthermore, we actually obtain a more general results, which the VWD threshold for the HMP can also be derived, please refer to Lemma \ref{lemm:mix0} for details. 

Furthermore, Theorem \ref{theo:three} establishes that the harmonic mean $M_{-1}$, the Cauchy combination test $M_{\tan}$, and the proposed method $M_{\rm{ptan}}$ share similar VAD and VWD thresholds. Actually, \cite{Chen2020TradeoffBV} also explored the asymptotic equivalence between the HMP and the CCT under arbitrary dependence structures. However, their established equivalence is limited to $\alpha \to 0^+$, while we remove this limitation by employing novel techniques.

\begin{theorem}\label{theo:three}
(i) Let $a_{\mathcal{H}}(\alpha)$, $a_{\mathcal{C}}(\alpha)$ and $a_{\mathcal{P}}(\alpha)$ be the VAD thresholds of $M_{-1}$, $M_{\tan}$ and the proposed $M_{\rm{ptan}}$, respectively. Let $\alpha \in (0, 1/2)$, as $K\to \infty$, 
\begin{align}
a_{\mathcal{P}}(\alpha)
\sim  a_{\mathcal{H}}(\alpha)
\sim  a_{\mathcal{C}}(\alpha)
\sim \frac{\alpha}{\log(K)}.
\end{align}
(ii) Let $b_{\mathcal{H}}(\alpha)$, $b_{\mathcal{C}}(\alpha)$ and $b_{\mathcal{P}}(\alpha)$ be the VWD thresholds of $M_{-1}$, $M_{\tan}$ and the proposed $M_{\rm{ptan}}$, respectively. Then, as $\alpha\to 0^+$ and $K\to\infty$,
\begin{align} \label{b_alpha}
b_{\mathcal{P}}(\alpha)
\sim  b_{\mathcal{H}}(\alpha)
\sim  b_{\mathcal{C}}(\alpha)  \sim \alpha.
\end{align}
\end{theorem}

Combining Theorem \ref{theo:r=-1} and Theorem \ref{theo:three}, $M_{\text{ptan}}$ shares the same sharp critical region for $M_{-1}$ and $M_{\text{tan}}$ under arbitrary dependence with any $\alpha \in (0, 1/2)$ and weak dependence with $\alpha \to 0^+$, i.e., the PCCT, the HMP and the CCT are asymptotically equivalent when $K \to \infty$ in those scenarios. 
An intuitive explanation is that, the discrepancy of both Cauchy related methods with the HMP can be controlled by an infinitesimal term as shown by Taylor formula that $1-2\arctan(x)/\pi=2/(\pi x)+o(1/x)$ as $x\to\infty$. 
Furthermore, they can all be categorized into the the regularly varying distributions, possessing properties that the null distributions are robust as the dependence structure varies in an asymptotic, but practical sense \citep{Fang2021HeavyTailedDF}. 
Up to this point, this section focuses on the relationship of the statistics and thresholds under weak dependence assumption and in the absence of any assumptions, uncovering the asymptotic equivalence of these methods under certain conditions. Although they exhibit asymptotic equivalence, their performances differ significantly in non-asymptotic scenarios, which will be demonstrated in simulations.
According to the conclusion derived from equation (\ref{b_alpha}), we presented the numerical values of $\alpha/b_{\phi}(\alpha)$ for the PCCT and HMP in Table \ref{table:bf} in the Supplementary Materials.
Notably, it is straightforward to see that $\alpha/b_{\phi}(\alpha)$ for the CCT consistently equals 1, regardless of the $\alpha$. 
Table \ref{table:bf} also reveals that the values of $a_{\phi}(\alpha)$ across the three methods are remarkably similar, and $\alpha/b_{\phi}(\alpha)$ for the PCCT and HMP tends to 1 as $\alpha \to 0^+$.

At the end of this section, we demonstrate the PCCT appears to enjoy advantages over the CCT and the Bonferroni method in terms of power. 
As mentioned earlier, based on Theorem \ref{theorem:vsd} and Theorem 1 in \cite{cct}, the power of PCCT outperforms than that of CCT consistently. Now this matter is presented formally in the Theorem \ref{theo:power} (i). Moreover, Theorem \ref{theo:power} (ii) shows that the proposed method is more efficient than the Bonferroni method as $\alpha \to 0^+$ and  $K\to \infty$. 


\begin{theorem}\label{theo:power} 
(i) Suppose conditions in Theorem \ref{theorem:vsd} hold. 
We have
\begin{align}
     {\rm pr}(T_{\rm{PCCT}}>t_{\alpha/2})>{\rm pr}(T_{\rm{CCT}}>t_{\alpha}).    
\end{align}
(ii) As $\alpha \to 0^+$ and  $K\to \infty$, under the alternative hypothesis, we have
\begin{align}
{\rm pr}\left(M_{\rm ptan}(p_1,\ldots,p_K) < b_{\mathcal{P}}(\alpha)\right) \geqslant
{\rm pr}\left(Kp_{(1)}<\alpha\right).
\end{align}
\end{theorem}

By Theorem \ref{theo:power} (i), it looks like the powers of PCCT enjoy advantages over the CCT. However, we notice that the power comparison is based on Theorem \ref{theorem:vsd}, which involves implicit condition $\alpha\to0^+$. Both Theorem \ref{theorem:vsd} and Theorem 1 in \cite{cct} approximate the tail probability when $\alpha\to0^+$, but for a fixed $\alpha$, the precision of the approximations may be different. It means the approximations differ in the extent to which they control sizes. Table \ref{table1} also shows that sizes for PCCT are harder to control. Therefore, it is meaningless to compare power since the PCCT encounters more serious type I error inflation. On the other side, this phenomenon necessitates developing testing with uniform type I error control under the same conditions such as VWD and VAD.

\section{Numerical Analysis}\label{sec:simulation}
\subsection{Performance evaluation under  null hypothesis}
In this subsection, we perform simulations to evaluate the validity of the proposed positive Cauchy combination testing (PCCT). Since a hand of theoretical properties among the statistics have been established in Section \ref{sec:property}, we compare its empirical performance with that of the Bonferroni method (BM), the harmonic mean $p$-value (HMP), and the Cauchy combination test (CCT). The VAD thresholds and the VWD thresholds are both used for the CCT, HMP and the proposed PCCT. 
The BM method rejects the null hypothesis when $Kp_{(1)}$ is less than the significance level $\alpha$.

We consider the following two settings for $\BX = (X_1, \ldots, X_K)$:
\begin{itemize}
    \item[]
    Case 1:  (\textit{Short-term dependence structure}) $\BX=(X_1, \dots, X_K)$ follows a multivariate normal distribution $N(\boldsymbol{\mu},\boldsymbol \Sigma_1)$.
    
    

    \item[] Case 2:  (\textit{Long-term dependence structure}) $\BX$ follows a multivariate normal distribution $N(\boldsymbol{\mu}, \boldsymbol \Sigma_2)$.
\end{itemize}
Here $\boldsymbol \Sigma_1 = (\rho_{ij})_{K\times K}$, where $\rho_{ij} = \rho^{\mid i-j\mid }$ representing auto-regression structure, 
while $\boldsymbol \Sigma_2 = (1-\rho) \boldsymbol I_K + \rho \boldsymbol J_K$, where $\boldsymbol I_K$ denotes the $K$-dimensional identity matrix and $\boldsymbol J_K$ represents a $K$-dimensional matrix with all elements equal to $1$. 
Our aim is to test $\boldsymbol H_0: \boldsymbol{\mu} = {\bf 0}$ versus $\boldsymbol H_1: \boldsymbol{\mu} > {\bf 0}$. 
Then $T_{\rm{CCT}}$ and $T_{\rm{PCCT}}$ can be characterized by combining $p_i = 1- \Phi( X_i)$ with \eqref{eq01} and \eqref{eq02}, respectively. 
Considering $K = 10^2, 10^3, \ldots, 10^6$ and $\rho = 0, 0.1, 0.2, 0.7$ to investigate the impact of the number of multiple tests $K$ and the correlation parameter $\rho$ on the results. The empirical sizes for Case 1 and Case 2 are calculated through $10^4$ repetitions with significance level $\alpha = 0.05$.

Table \ref{table:Nullcase1} displays the empirical sizes for Case 1 and implies that the VWD thresholds for CCT, HMP and PCCT can almost control the type I error with any $\rho$ under auto-regression structure, where weak dependence in Assumption \ref{ass2} holds. Thus the VWD thresholds are valid in such a case, which is in accordance with the theoretical results. The sizes of the null $\boldsymbol H_0$ for Case 2 are displayed in Table \ref{table:Nullcase4}, which indicates that the four methods with the VAD thresholds can control the type I error in the all cases. With the VWD thresholds, however, the three methods fail even with a small $\rho \neq 0$ in this case, where the long-term dependence covariance setting may violate Assumption \ref{ass2}. Therefore, in such scenorios, the VAD thresholds are preferred due to their size guarantee. 





\begin{table}[ht]
\caption{Empirical sizes for Case 1 where $\rho$ and $K$ values vary}
\centering
\small
\begin{tabular}{ccccccccc}
\toprule
\multirow{2}*{$\rho$} & \multirow{2}*{$K$} & & \multicolumn{3}{c}{VAD thresholds} & \multicolumn{3}{c}{VWD thresholds}\\
\cmidrule(lr){4-6} \cmidrule(lr){7-9}
&  & BM & CCT & HMP & PCCT & CCT & HMP & PCCT \\ 
\midrule
0& $10^2$  & 0.0474 & 0.0070 & 0.0077 & 0.0076 & 0.0485 & 0.0491 & 0.0490 \\ 
& $10^3$ & 0.0496 & 0.0060 & 0.0064 & 0.0064 & 0.0512 & 0.0504 & 0.0503 \\ 
& $10^4$ & 0.0491 & 0.0039 & 0.0040 & 0.0040 & 0.0501 & 0.0490 & 0.0490 \\ 
& $10^5$ & 0.0476 & 0.0039 & 0.0039 & 0.0039 & 0.0492 & 0.0496 & 0.0496 \\
& $10^6$ & 0.0451 & 0.0027 & 0.0028 & 0.0028 & 0.0444 & 0.0445 & 0.0452 \\ 
\midrule

0.1 & $10^2$ & 0.0462 & 0.0062 & 0.0065 & 0.0065 & 0.0495 & 0.0486 & 0.0486 \\ 
& $10^3$ & 0.0448 & 0.0048 & 0.0051 & 0.0051 & 0.0459 & 0.0464 & 0.0463 \\ 
& $10^4$ & 0.0452 & 0.0045 & 0.0045 & 0.0045 & 0.0441 & 0.0449 & 0.0449 \\ 
& $10^5$ & 0.0488 & 0.0030 & 0.0031 & 0.0031 & 0.0492 & 0.0500 & 0.0500 \\ 
& $10^6$ & 0.0436 & 0.0018 & 0.0020 & 0.0020 & 0.0439 & 0.0450 & 0.0460 \\ 
\midrule
0.2 & $10^2$  & 0.0508 & 0.0072 & 0.0079 & 0.0079 & 0.0544 & 0.0533 & 0.0533 \\ 
& $10^3$ & 0.0474 & 0.0054 & 0.0054 & 0.0054 & 0.0475 & 0.0489 & 0.0488 \\ 
& $10^4$ & 0.0490 & 0.0043 & 0.0048 & 0.0048 & 0.0498 & 0.0490 & 0.0490 \\ 
& $10^5$ & 0.0506 & 0.0031 & 0.0032 & 0.0032 & 0.0514 & 0.0512 & 0.0512 \\ 
& $10^6$ & 0.0479 & 0.0029 & 0.0032 & 0.0032 & 0.0460 & 0.0471 & 0.0478 \\ 
\midrule
0.7 & $10^2$ & 0.0414 & 0.0088 & 0.0090 & 0.0090 & 0.0594 & 0.0539 & 0.0541 \\ 
& $10^3$ & 0.0430 & 0.0053 & 0.0057 & 0.0057 & 0.0537 & 0.0510 & 0.0508 \\ 
& $10^4$ & 0.0448 & 0.0033 & 0.0035 & 0.0035 & 0.0540 & 0.0501 & 0.0501 \\ 
& $10^5$ & 0.0440 & 0.0038 & 0.0039 & 0.0039 & 0.0486 & 0.0468 & 0.0468 \\
& $10^6$ & 0.0483 & 0.0025 & 0.0027 & 0.0027 & 0.0515 & 0.0504 & 0.0514 \\ 
\bottomrule
\end{tabular}
\label{table:Nullcase1}
\end{table}

\begin{table}[ht]
\caption{Empirical sizes for Case 2 where $\rho$ and $K$ values vary}
\centering
\small
\begin{tabular}{ccccccccc}
  \toprule
  \multirow{2}*{$\rho$} & \multirow{2}*{$K$} & & \multicolumn{3}{c}{VAD thresholds} & \multicolumn{3}{c}{VWD thresholds}\\
 \cmidrule(lr){4-6} \cmidrule(lr){7-9}
 &  & BM & CCT & HMP & PCCT & CCT & HMP & PCCT \\ 
  \midrule
0  & $10^2$ & 0.0466 & 0.0062 & 0.0064 & 0.0063 & 0.0468 & 0.0468 & 0.0469 \\ 
  & $10^3$ & 0.0489 & 0.0049 & 0.0060 & 0.0060 & 0.0503 & 0.0517 & 0.0516 \\ 
  & $10^4$ & 0.0515 & 0.0054 & 0.0056 & 0.0056 & 0.0521 & 0.0527 & 0.0527 \\ 
  & $10^5$ & 0.0479 & 0.0029 & 0.0029 & 0.0029 & 0.0478 & 0.0478 & 0.0478 \\ 
  & $10^6$ & 0.0485 & 0.0022 & 0.0022 & 0.0022 & 0.0502 & 0.0486 & 0.0497 \\ 
  \midrule
0.1  & $10^2$ & 0.0432 & 0.0087 & 0.0087 & 0.0087 & 0.0757 & 0.0610 & 0.0616 \\ 
  & $10^3$ & 0.0475 & 0.0047 & 0.0048 & 0.0048 & 0.1010 & 0.0733 & 0.0735 \\ 
  & $10^4$ & 0.0457 & 0.0054 & 0.0055 & 0.0055 & 0.1352 & 0.0902 & 0.0905 \\ 
  & $10^5$ & 0.0377 & 0.0058 & 0.0059 & 0.0059 & 0.1583 & 0.0974 & 0.0977 \\ 
  & $10^6$ & 0.0380 & 0.0045 & 0.0045 & 0.0045 & 0.1754 & 0.1032 & 0.1057 \\ 
  \midrule 
0.2  & $10^2$ & 0.0404 & 0.0082 & 0.0083 & 0.0083 & 0.0852 & 0.0620 & 0.0622 \\
  & $10^3$ & 0.0409 & 0.0080 & 0.0080 & 0.0080 & 0.1116 & 0.0786 & 0.0790 \\ 
  & $10^4$ & 0.0348 & 0.0076 & 0.0076 & 0.0076 & 0.1350 & 0.0872 & 0.0886 \\ 
  & $10^5$ & 0.0300 & 0.0068 & 0.0069 & 0.0069 & 0.1452 & 0.0878 & 0.0881 \\ 
  & $10^6$ & 0.0285 & 0.0071 & 0.0071 & 0.0071 & 0.1536 & 0.0896 & 0.0908 \\ 
\midrule 
0.7  & $10^2$ & 0.0152 & 0.0090 & 0.0090 & 0.0090 & 0.0686 & 0.0500 & 0.0504 \\ 
  & $10^3$ & 0.0072 & 0.0069 & 0.0069 & 0.0069 & 0.0670 & 0.0472 & 0.0475 \\ 
  & $10^4$ & 0.0035 & 0.0071 & 0.0071 & 0.0071 & 0.0661 & 0.0408 & 0.0411 \\ 
  & $10^5$ & 0.0030 & 0.0057 & 0.0057 & 0.0057 & 0.0738 & 0.0450 & 0.0455 \\ 
  & $10^6$ & 0.0005 & 0.0042 & 0.0042 & 0.0042 & 0.0716 & 0.0397 & 0.0403 \\ 
   \bottomrule
\end{tabular}
\label{table:Nullcase4}
\end{table}

\subsection{Statistical power}
In this subsection, we compare the statistical power of BM, CCT, HMP and the proposed PCCT under Case 1 and Case 2 in the former subsection for convenience. 
As there is no evidence against the weak dependence assumption for Case 1, we calculate the power with VWD thresholds, whereas VAD thresholds for the Case 2. For fixed $K = 10^3$, the powers are calculated through $10^4$ repetitions. We consider the sparse signals and the dense signals of $\boldsymbol{\mu}$ as follows. 

\begin{itemize}
    \item (\textit{Sparse signals}): Let the mean vector $\boldsymbol{\mu} = (\mu_{1},\ldots,\mu_{K})$ with $|\mu_{i}| = c_{0}$ for $1\le i\le 0.05 K$ or $ 0.5K+1 \le i\le 0.55K+1$, where $c_0=0.2, 0.4, \ldots, 2.0$ to denote the signal strength. Let $\mu_{i}=0$ for the other $i$.


    \item (\textit{Dense signals}): Let the mean vector $\boldsymbol{\mu} = (\mu_{1},\ldots,\mu_{K})$ with $|\mu_{i}| = c_{0}$ for $1\le i\le K$, where $c_0=0.1,0.2,\ldots,1.0$ to denote the signal strength.    

\end{itemize}

We next consider two types of signal directions, the one is to let all the means are positive, while the other is to let the first half be positive and the second half be negative. The latter setting for negative signals aims for generating $p$-values close to 1 to simulate the negative penalty for the CCT as discussed in Section \ref{sec2}.


\begin{figure}
\begin{center}
\includegraphics[width=14.5cm]{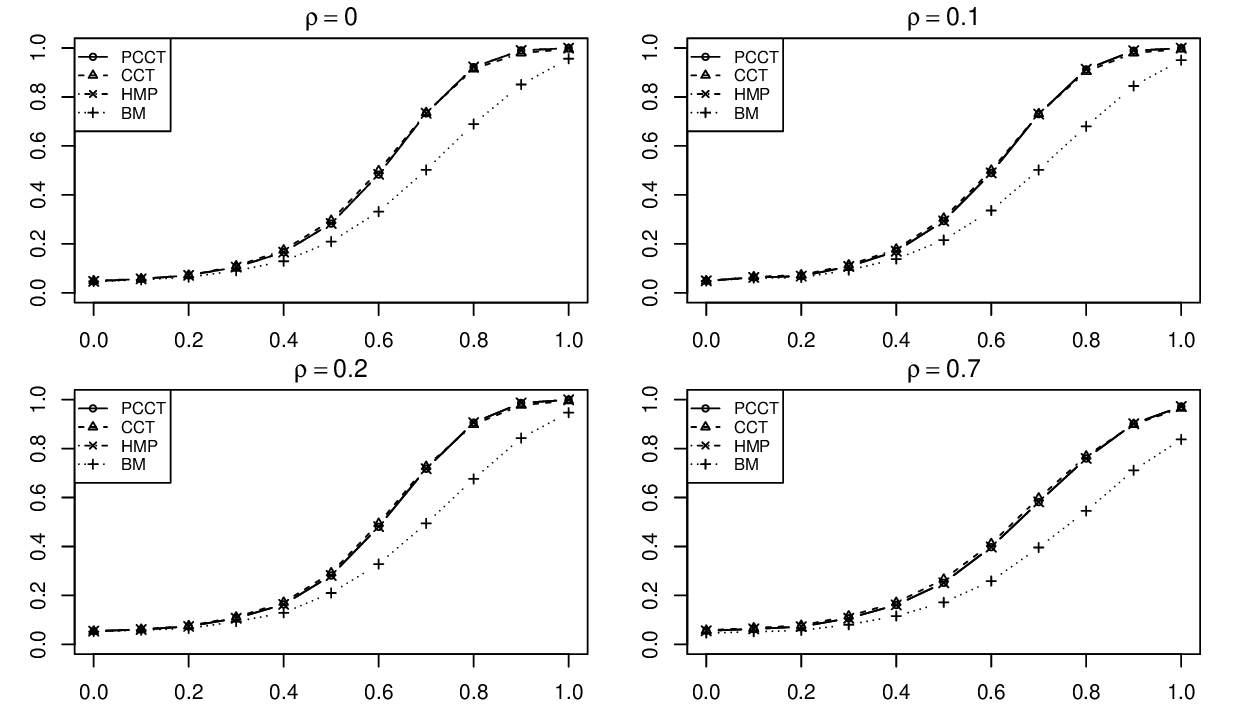}
\caption{Power curves of BM, CCT, HMP and PCCT against the sparse signals without negative penalties under $\rho = 0, 0.1, 0.2, 0.7$ and $K = 10^3$ for Case 1. }\label{figure:powercase1_1}
~\\
\includegraphics[width=14.5cm]{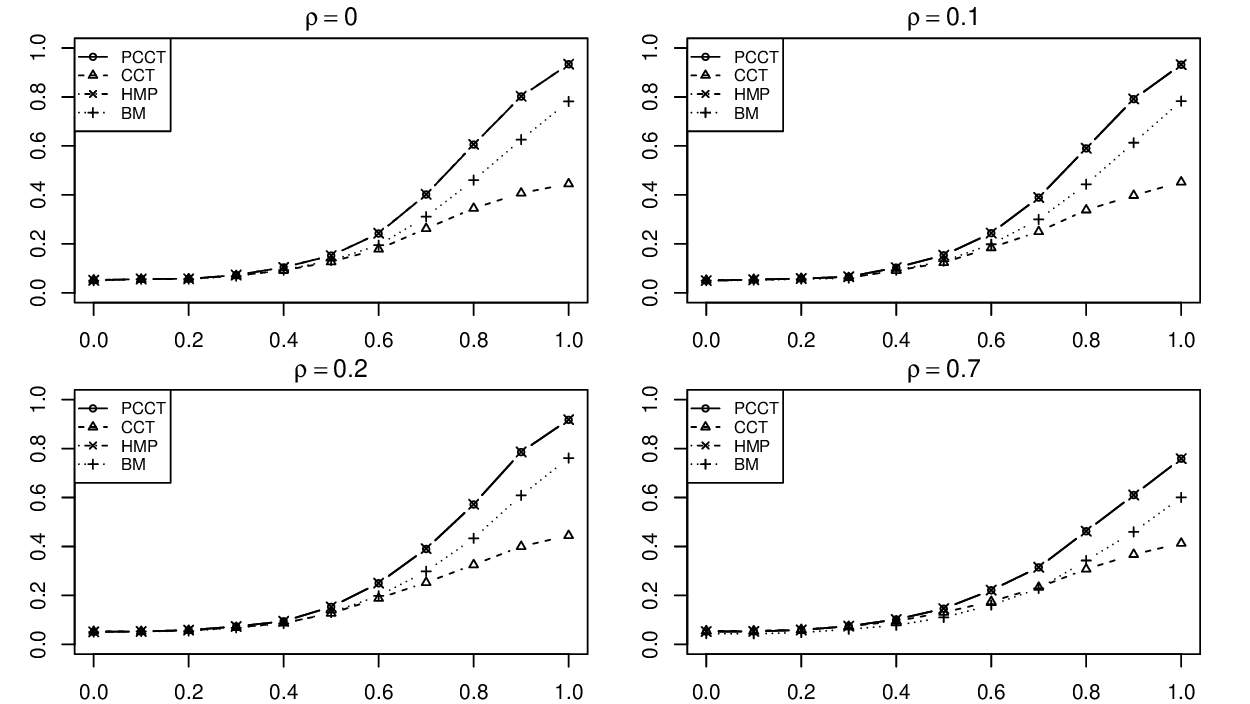}
\caption{Power curves of BM, CCT, HMP and PCCT against the sparse signals with negative penalties under $\rho = 0, 0.1, 0.2, 0.7$ and $K = 10^3$ for Case 1. }\label{figure:powercase1_2}
\end{center}
\end{figure}

\begin{figure}
\begin{center}
\includegraphics[width=14.5cm]{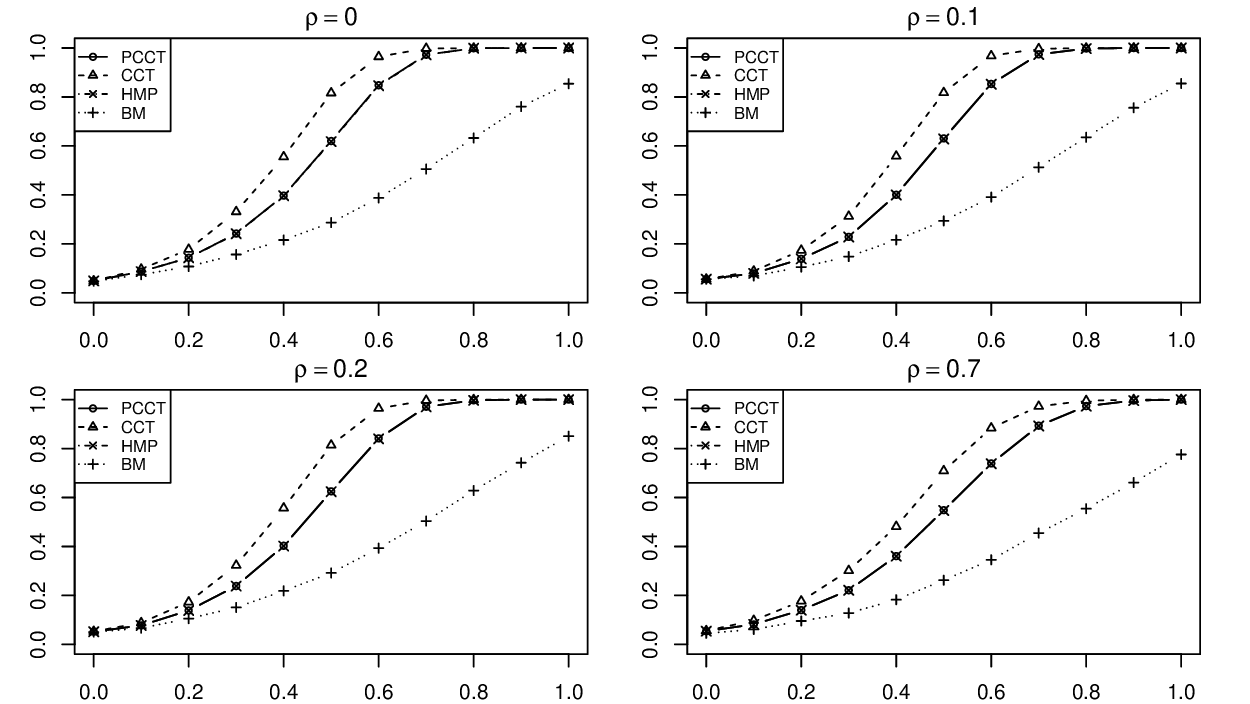}
\caption{Power curves of BM, CCT, HMP and PCCT against the dense signals without negative penalties under $\rho = 0, 0.1, 0.2, 0.7$ and $K = 10^3$ for Case 1. }\label{figure:dpowercase1_1}
~\\
\includegraphics[width=14.5cm]{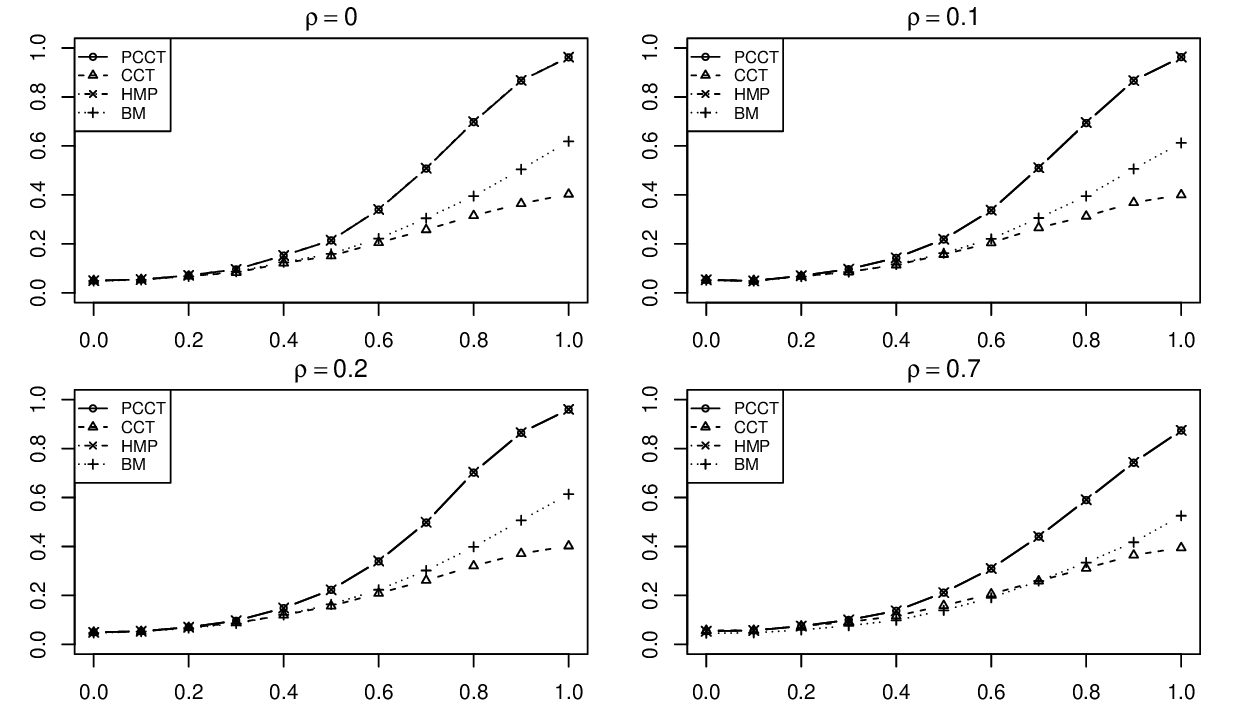}
\caption{Power curves of BM, CCT, HMP and PCCT against the dense signals with negative penalties under $\rho = 0, 0.1, 0.2, 0.7$ and $K = 10^3$ for Case 1. }\label{figure:dpowercase1_2}
\end{center}
\end{figure}

For Case 1, Figures \ref{figure:powercase1_1}-\ref{figure:dpowercase1_2} show the power curves of the BM, HMP, CCT and PCCT under $\rho = 0, 0.1, 0.2, 0.7$. From Fig. \ref{figure:powercase1_1} and Fig. \ref{figure:dpowercase1_1}, it can be seen that the PCCT performs almost identical to the CCT for sparse signals without negative penalties, while the CCT generates slightly higher power for dense signals. 
As shown in Fig. \ref{figure:powercase1_2} and Fig. \ref{figure:dpowercase1_2}, the PCCT enjoys the superior performance among four methods for observations with negative penalties. Since there exist some $p$-values near to 1, the CCT can not work well under these settings. Meanwhile, the HMP has a similar performance to the proposed method, which can be seen from both empirical sizes and power analysis for Case 1.

For Case 2, 
the power curves for BM are omitted since sizes for BM are far from those for CCT, HMP and PCCT with VAD thresholds a lot, as shown in Table \ref{table:Nullcase4}. 
For the space limitation, their corresponding powers are displayed in Fig. \ref{figure:powercase4_1}-\ref{figure:dpowercase4_2} in the Supplementary Materials, which indicates their powers share similar relationships like Case 1. 


In summary, the PCCT stands out among all the methods with negative penalties, which addresses our initial concerns effectively. When comes to the sparse signals without negative penalties, the PCCT also remains almost the same power with the CCT, while the CCT performs better with the dense signals. 
However, the combination test is not suitable for such dense signals in practical scenarios. 
It is because the effects with dense signals may be evident and thus the global testing may be unnecessary. As a result, the combination test, which is designed to preform global testing to combine weak and sparse signals, seems to be meaningless when testing dense signals. 
The above discussion aims to elucidate that the combination methods are appropriate for weak and sparse signals, where the PCCT enjoys advantages over others.

\subsection{Analysis of real data}
In this subsection, we apply the proposed method to a genome-wide association studies (GWAS) of neuroticism (\cite{2016data}). 
Firstly, we perform the single-SNP analysis with neuroticism containing 6,524,432 genetic variants (SNPs) across 179,811 individuals. By the Bonferroni method, we can find 3,199 SNPs significant at $\alpha = 0.05$. After excluding these 3,199 SNPs, we combine the individual $p$-values of the remaining 6,521,233 SNPs with the proposed method, and the test statistics is $8.5 \times 10^{-5} < a_{\mathcal{P}}(0.05) < b_{\mathcal{P}}(0.05)$, where $a_{\mathcal{P}}$ and $b_{\mathcal{P}}$ correspond the PCCT using VAD threshold and VWD threshold, respectively. Therefore, there still exists genetic information in the remaining SNPs.

Next, we combine the $p$-values of SNPs within contiguous regions for $K = 10, 10^2, \ldots, 10^5$ and the entire chromosomes (Chr). Table \ref{table:app1} shows the number of statistically significant regions for the BM, HMP, CCT and PCCT, where the second column displays the size of these regions. Considering the VWD thresholds, it can be seen that BM performs conservatively compared with PCCT and HMP. Moreover, PCCT with the VWD thresholds can detect more significant regions than HMP for a small $K = 10, 10^2$. As for the VAD thresholds, the numbers of significant regions for PCCT and HMP are almost identical, and the two methods seem to be more conservative than BM. 
For the CCT, while it is capable of identifying thousands of regions when $K$ is small, the number of significant regions sharply declines as $K$ increases due to the large negative penalties. The significance is almost worn away when $K$ reaches $10^5$ or tests are conducted across entire chromosomes.
In summary, our proposed PCCT with the VWD thresholds can detect more possible effective SNP regions, while the other methods may miss those signals and discoveries.

  

\begin{table}
\caption{The number of significant regions for $K = 10, 10^2, \ldots, 10^5$ and entire chromosomes (Chr)}
\centering
\begin{tabular}{rrrrrrrrr}
\toprule
\multirow{2}*{$K$} & \multirow{2}*{Regions} & & \multicolumn{3}{c}{VAD thresholds} & \multicolumn{3}{c}{VWD thresholds}\\
 \cmidrule(lr){4-6} \cmidrule(lr){7-9}
&  & BM & CCT  & HMP & PCCT  & CCT  & HMP  & PCCT  \\ 
\midrule
 10 & 652443 & 40650 &  25322  & 25794 & 25762 & 68896 & 62808 &  63421\\ 
 $10^2$ & 65244 & 2929 & 2295 & 2335 & 2333 & 7903 & 7177 & 7203 \\ 
 $10^3$ & 6524 & 354 & 282 & 295 & 295 & 1022 & 966 & 965 \\ 
 $10^4$ & 652 & 77 & 36 & 56 & 56 & 156 & 203 & 204 \\ 
 $10^5$ & 65 & 17 & 1 & 16 & 16 & 2 & 42 & 42 \\ 
 Chr & 22 & 10 & 0 & 7 & 7 & 0 & 19 & 19 \\ 
  \bottomrule
\end{tabular}
\label{table:app1}
\end{table}

\section{Conclusion}\label{sec:conclusion}
Cauchy combination test has garnered significant recognition due to its simplicity and powerful effectiveness. \cite{cct} derived the approximation of tail probabilities as the significance level $\alpha\to 0^+$, however, the method may loss substantial power with some $p$-values near to 1. In addition, Cauchy combination test may fail to control the type I error for a given significance level.
In this paper, we propose Positive Cauchy combination test, which can be less affected by extremely large $p$-values. The limit distribution under weak dependence has been derived to promise the method can work well for any significance level $\alpha$. Further, we provide the threshold for the proposed method without any additional assumptions, however, the proposed method with this threshold seems to be conservative. Therefore, it is interesting to extend the method to a more general dependence structure and maintain effective power. We leave these topics for possible future research.









\section*{Acknowledgement}
The authors gratefully acknowledge Beijing Natural Science Foundation (No Z200001), National Natural Science Foundation of China (No 11971478), and by Public Health \& Disease Control and Prevention, Major Innovation \& Planning Interdisciplinary Platform for the ``Double-First Class'' Initiative, Renmin University of China (No. 2021PDPC).

\section*{Supplementary material}
\label{SM}
The supplementary material includes the Appendix which gives additional  numerical results and all proofs.


\bibliographystyle{apalike} 
\bibliography{reference}

\newpage
\appendix

\bigskip\bigskip

{\fontsize{19.5pt}{14pt}\bf  Supplementary material for ``Effective Positive Cauchy Combination Test"}

\bigskip\bigskip\bigskip\bigskip
\setcounter{table}{0}
\renewcommand{\thetable}{A\arabic{table}}

\setcounter{figure}{0}
\renewcommand{\thefigure}{A\arabic{figure}}

The Supplementary Material includes additional  numerical results and all proofs.
Appendix \ref{asec:simulation} presents 
powers for Case 2 and the numerical values of thresholds. 
Appendix \ref{asec:proof} collects all technical lemmas and proofs of the main theorems.

\section{The supplemental numerical results}\label{asec:simulation}

\subsection{The numerical values of thresholds}

In this subsection, we report numerical results of the VAD threshold and the VWD threshold with $K = 10, 10^2, 10^3, 10^4, 10^5, 10^6, 10^7, 10^8$. 
For comparison, we also consider the harmonic mean $p$-value (HMP) in \cite{2019HMP} and the CCT in \cite{cct}. Note that $a_{\mathcal{H}}(\alpha) \to \alpha/\log(K)$ as $K\to\infty$(\cite{vovk2020, Chen2020TradeoffBV}), we calculate numerical values of $\alpha/\{\log(K)a_{\phi}(\alpha)\}$ for the three methods with the significance level $\alpha = 0.05, 0.01, 0.001$. Table \ref{table:af} reports the above results and shows the three methods share similar critical region as Theorem \ref{theo:three} (i) implies. We also notice that $\alpha/\{\log(K)a_{\mathcal{P}}(\alpha)\}>1.22$ for a very large $K = 10^8$, which implies the rate of convergence is very slow.
Table \ref{table:bf} displays the numerical values of $\alpha/b_\phi(\alpha)$ for the harmonic mean and the proposed method with the significance level $\alpha = 0.05, 0.01, 0.001$. For a small significance level $\alpha = 0.001$, the VWD thresholds for harmonic mean and the proposed method are close to 1 as Theorem \ref{theo:three} (ii) shows.

\begin{figure}
\begin{center}
\includegraphics[width=14.5cm]{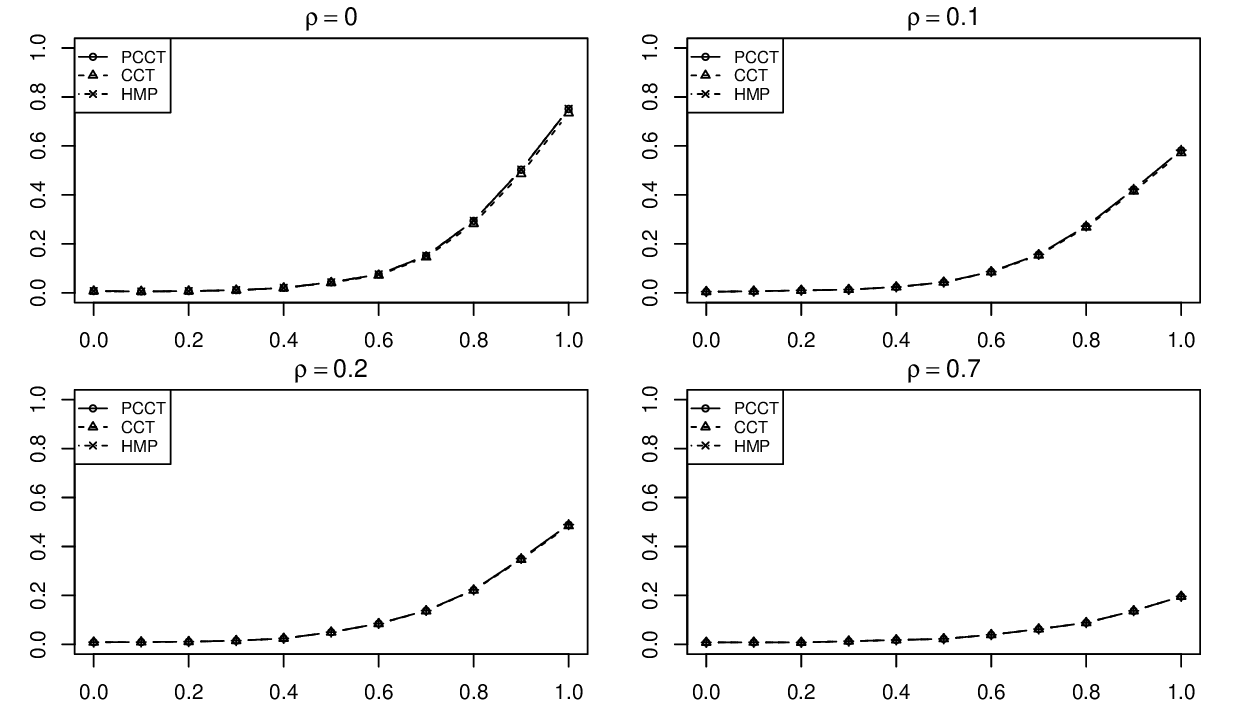}
\caption{Power curves of CCT, HMP  and PCCT against the sparse signals without negative penalties under $\rho = 0, 0.1, 0.2, 0.7$ and $K = 10^3$ for Case 2. }\label{figure:powercase4_1}
~\\
\includegraphics[width=14.5cm]{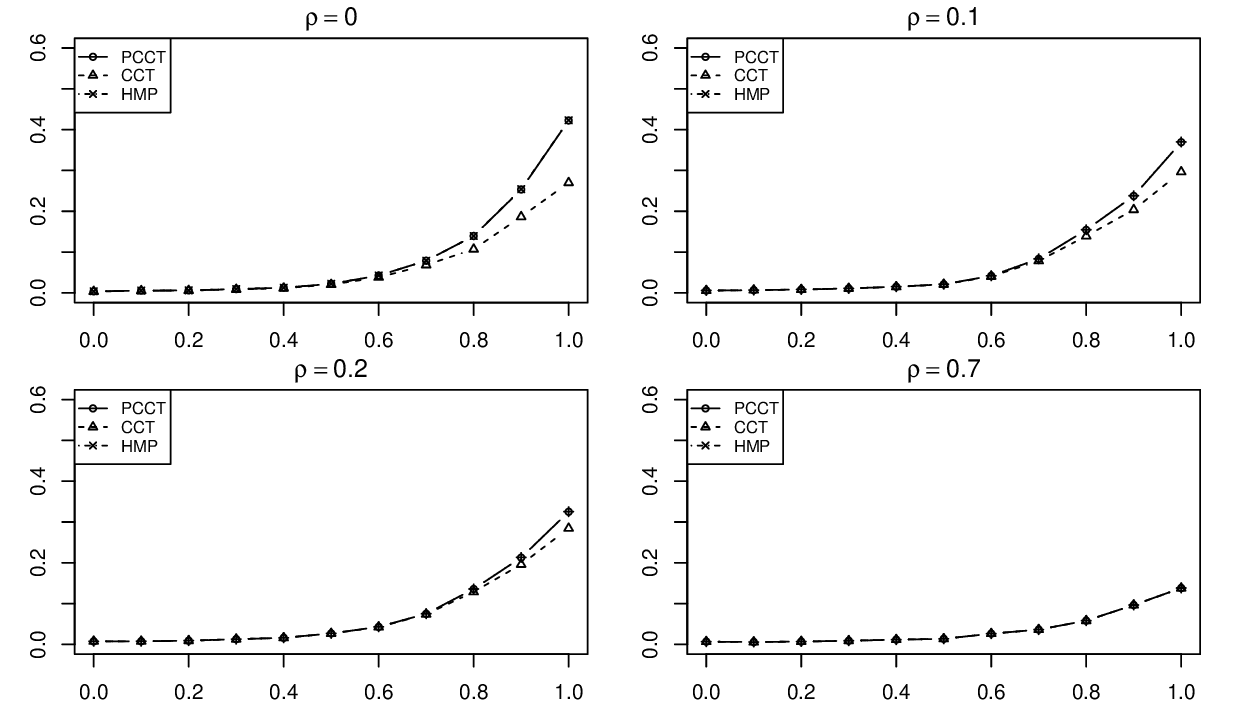}
\caption{Power curves of CCT, HMP and PCCT against the sparse signals with negative penalties under $\rho = 0, 0.1, 0.2, 0.7$ and $K = 10^3$ for Case 2. }\label{figure:powercase4_2}
\end{center}
\end{figure}

\begin{figure}
\begin{center}
\includegraphics[width=14.5cm]{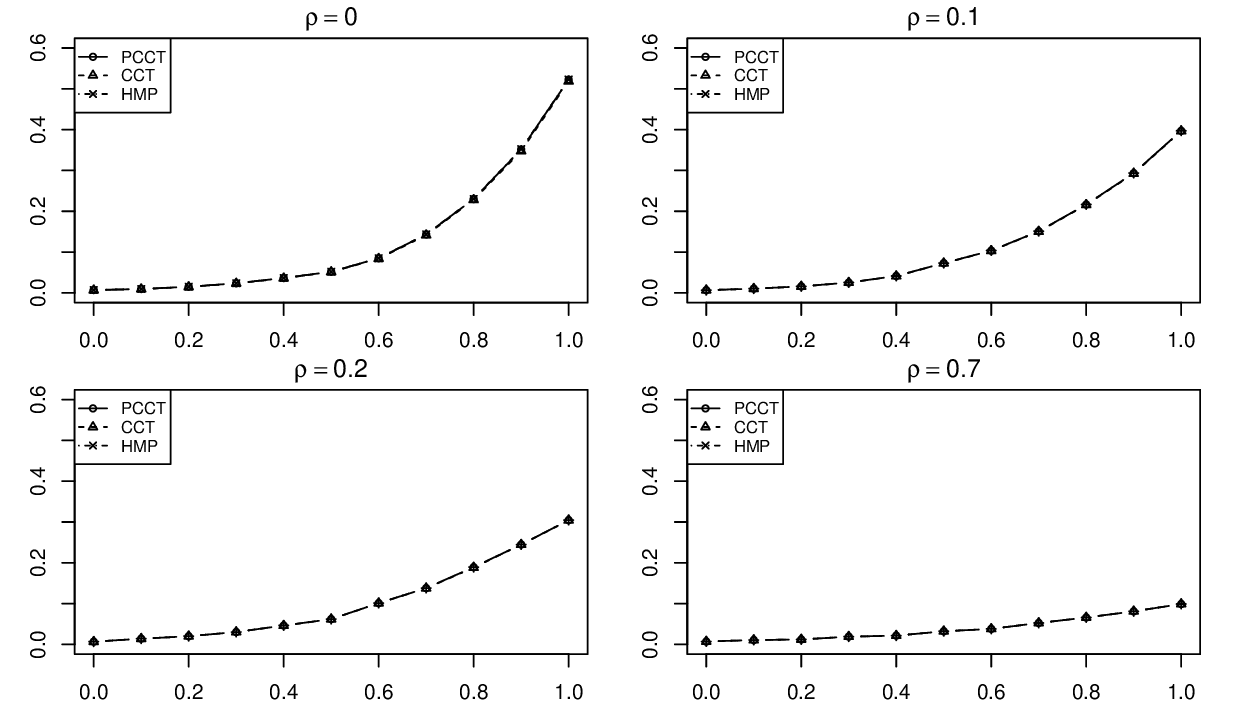}
\caption{Power curves of CCT, HMP and PCCT against the dense signals without negative penalties under $\rho = 0, 0.1, 0.2, 0.7$ and $K = 10^3$ for Case 2. }\label{figure:dpowercase4_1}
~\\
\includegraphics[width=14.5cm]{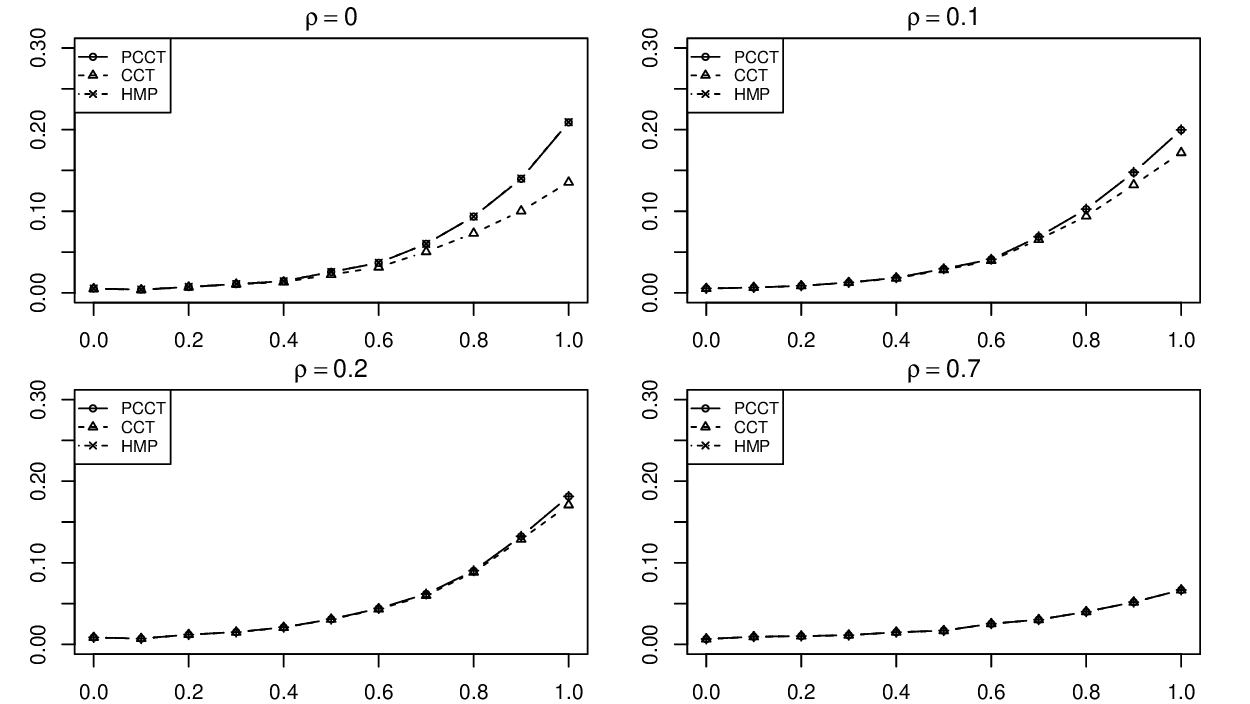}
\caption{Power curves of CCT, HMP  and PCCT against the dense signals with negative penalties under $\rho = 0, 0.1, 0.2, 0.7$ and $K = 10^3$ for Case 2. }\label{figure:dpowercase4_2}
\end{center}
\end{figure}

\begin{table}[ht]
\caption{Numerical values of $\alpha/\{\log(K)a_{\phi}(\alpha)\}$ for the HMP, the CCT and the proposed PCCT with the significance level $\alpha = 0.1, 0.05, 0.01$}
\centering
\small
\begin{tabular}{cccccccccc}
  \toprule  
\multirow{2}*{$K$} &  \multicolumn{3}{c}{$\alpha = 0.1$} 
    &  \multicolumn{3}{c}{$\alpha = 0.05$}
    &  \multicolumn{3}{c}{$\alpha = 0.01$} \\
    \cmidrule(lr){2-4} \cmidrule(lr){5-7}\cmidrule(lr){8-10}
 & CCT & HMP & PCCT & CCT & HMP & PCCT & CCT & HMP & PCCT \\ 
 \midrule
10 & 1.9781 & 1.9803 & 1.9798 & 1.9798 & 1.9803 & 1.9802 & 1.9803 & 1.9803 & 1.9803 \\ 
  $10^2$ & 1.6176 & 1.6196 & 1.6191 & 1.6191 & 1.6196 & 1.6195 & 1.6196 & 1.6196 & 1.6196 \\ 
  $10^3$ & 1.4620 & 1.4637 & 1.4633 & 1.4633 & 1.4637 & 1.4635 & 1.4636 & 1.4637 & 1.4637 \\ 
  $10^4$ & 1.3735 & 1.3748 & 1.3745 & 1.3745 & 1.3748 & 1.3748 & 1.3748 & 1.3748 & 1.3749 \\ 
  $10^5$ & 1.3157 & 1.3168 & 1.3166 & 1.3166 & 1.3168 & 1.3168 & 1.3169 & 1.3168 & 1.3169 \\ 
  $10^6$ & 1.2747 & 1.2757 & 1.2755 & 1.2755 & 1.2757 & 1.2756 & 1.2757 & 1.2757 & 1.2757 \\ 
  $10^7$ & 1.2440 & 1.2448 & 1.2446 & 1.2446 & 1.2448 & 1.2448 & 1.2448 & 1.2448 & 1.2448 \\ 
  $10^8$ & 1.2200 & 1.2207 & 1.2206 & 1.2206 & 1.2207 & 1.2207 & 1.2207 & 1.2207 & 1.2208 \\ 
   \bottomrule
\end{tabular}
\label{table:af}
\end{table}

\begin{table}[ht]
\caption{Numerical values of $\alpha/b_\phi(\alpha)$ for the HMP and the PCCT with $\alpha = 0.05, 0.01, 0.001$}
\centering
\begin{tabular}{ccccccc}
  \toprule
\multirow{2}*{$K$}&  \multicolumn{2}{c}{$\alpha = 0.05$} 
    &  \multicolumn{2}{c}{$\alpha = 0.01$}
    &  \multicolumn{2}{c}{$\alpha = 0.001$} \\
\cmidrule(lr){2-3} \cmidrule(lr){4-5} \cmidrule(lr){6-7}
 & HMP & PCCT & HMP & PCCT & HMP & PCCT \\  
 \midrule
10 & 1.2595 & 1.2342 & 1.0690 & 1.0609 & 1.0092 & 1.0054 \\ 
  $10^2$ & 1.3743 & 1.3542 & 1.0919 & 1.0877 & 1.0115 & 1.0111 \\ 
  $10^3$ & 1.4968 & 1.4787 & 1.1164 & 1.1126 & 1.0140 & 1.0136 \\ 
  $10^4$ & 1.6149 & 1.5940 & 1.1400 & 1.1357 & 1.0163 & 1.0159 \\ 
  $10^5$ & 1.7273 & 1.7066 & 1.1625 & 1.1582 & 1.0186 & 1.0181 \\ 
  $10^6$ & 1.8452 & 1.8035 & 1.1861 & 1.1776 & 1.0209 & 1.0201 \\ 
  $10^7$ & 1.9407 & 1.9338 & 1.2052 & 1.2037 & 1.0228 & 1.0227 \\ 
  $10^8$ & 2.0712 & 2.0542 & 1.2313 & 1.2278 & 1.0255 & 1.0251 \\ 
  \bottomrule
\end{tabular}
\label{table:bf}
\end{table}

\bigskip\bigskip

\section{Technical proofs}\label{asec:proof}

\subsection{Proof of Theorem \ref{theorem:vsd}}

\begin{proof}
~\cite{Fang2021HeavyTailedDF} discussed the tail probabilities of partial sums of regular variable random variables under the assumption of binary normality. For the PCCT statistic, we also have the following lemma.
\begin{lemma}\label{lemm:kt}
Under the assumption of binary normality, as $t\to \infty$, we have
\begin{align}\label{eq:kt}
{\rm pr}\left(KT_{\rm PCCT} > t\right) \sim K {\rm pr}(W_{\mathcal{C}} > t), 
\end{align}
where the distribution function of $W_{\mathcal{P}}$ is $F_{\mathcal{P}}(u) = 2\arctan(u)/\pi$.
\end{lemma}
Actually, when $p$-values are independently and identically distributed, \eqref{eq:kt} still holds. For fixed $K$, we have
\begin{align}
{\rm pr}\left(T_{\rm PCCT} > t\right) \sim K {\rm pr}(W_{\mathcal{C}} > Kt) = K(1-F_{\mathcal{P}}(Kt)) \sim {\rm pr}(W_{\mathcal{C}} > t) = 2{\rm pr}(W_{\mathcal{C}} > t).
\end{align}
which completes the proof.
\end{proof}




\subsection{Proof of Theorem \ref{th:weak}}

To discuss the asymptotic distribution of the proposed method under weak dependence, we introduce the stable random variable briefly.  
A stable random variable $Y\sim S(\alpha,\beta,\gamma,\delta)$ can be described by its characteristic function as
\begin{align}
E\exp\{iYt\} = \exp\{i\delta t - \gamma^\alpha|t|^\alpha(1-i\beta \textrm{sign}(t)z(t,\alpha)\},\quad t\in \mathbb{R},
\end{align}
where $\delta$ is a location parameter, $\gamma>0$ is a scale parameter, 
$\alpha\in (0,2]$, $\beta\in[-1,1]$, and
\begin{align*}
z(t,\alpha) = \left\{
\begin{array}{ll}
     \tan\left(\frac{\pi\alpha}{2}\right) & \textrm{if }\alpha\not=1,\\
     -\frac{2}{\pi}\ln|t| & \textrm{if }\alpha = 1.
\end{array}
\right.
\end{align*}

\begin{lemma}\label{lemm:mix0}  Suppose Assumption \ref{ass1}-\ref{ass2} hold and $E(X_i) = 0, 1\le i\le K$. Let $T_\phi = \sum_{i=1}^{K}Y_i /K$, where $Y_i = \phi(\Phi(X_i))\sim W_\phi$ is regularly varying with index 1. The distribution function of $W_\phi$ is denoted by $F_\phi(x)$, and then the right tail of $W_\phi$ satisfies $1-F_\phi(x) \to c_\phi/x$ as $x\to \infty$ for fixed $c_\phi$.\\
(i) If  $W_\phi$ is symmetric, we have
\begin{align}
T_{\phi} \to S\left(1, 0, c_\phi\pi, 0\right)\qquad {\rm as}\ K\to \infty.
\end{align}
(ii) If  $W_\phi\ge 0$, we have
\begin{align}
T_{\phi} -  \Delta_\phi \to S\left(1, 1, \frac{c_\phi\pi}{2}, 0\right)\qquad {\rm as}\ K\to \infty,
\end{align}
where $\Delta_\phi = K\mathbb{E}\left(\sin\left(W_\phi/K\right)\right)$.
\end{lemma}



\begin{corollary}\label{coll:hmp} 
Suppose Assumption \ref{ass1}-\ref{ass2} hold and $E(X_i) = 0, 1\le i\le K$. Then we have
\begin{align}
T_{\rm HMP} -  \Delta_{\mathcal{H}} \to S\left(1, 1, \frac{\pi}{2}, 0\right)\qquad {\rm as}\ K\to \infty,
\end{align}
where $\Delta_{\mathcal{H}} = K\mathbb{E}\left(\sin\left(W_{\mathcal{H}}/K\right)\right)$ and the distribution function of $W_{\mathcal{H}}$ is $F_{\mathcal{H}}(x) = 1-1/x$.
\end{corollary}

{\bf Proof of Lemma \ref{lemm:mix0}: }

\begin{proof}
To prove Lemma \ref{lemm:mix0}, we need to verify the conditions (RV), (MX), (AC), (TB) and (CT) in \cite{2011dependent}. Since $W_\phi$ is a strictly stationary sequence and regularly varying with index $1$, the condition (RV) is satisfied. From Assumption \ref{ass2} and Lemma 3.8 in \cite{2011dependent}, the condition (MX) is satisfied. 
Let $c_{K, \phi}$ satisfy
\begin{align}
K{\rm pr}(|W_\phi|>c_{K,\phi}) \sim 1.
\end{align}
We note that 
\begin{align*}
&\lim_{h \to\infty}\limsup_{K\to\infty}K{\rm pr}\Big(\Big|\sum_{i=h}^{r_n}Y_i\Big|>c_{K,\phi},|X_1|>c_{K,\phi}\Big) \\
\le & \lim_{h \to\infty}\limsup_{K\to\infty}K{\rm pr}\Big(\Big|\sum_{i=h}^{r_n}Y_i\Big|>c_{K,\phi}\Big){\rm pr}\left(|X_1|>c_{K,\phi}\right) + \lim_{h \to\infty}\limsup_{K\to\infty}K\alpha_{h} \\
\le & 0.
\end{align*}
From Lemma 3.7 in \cite{2011dependent}, the condition (AC) is satisfied. 
Next, we discuss Lemma \ref{lemm:mix0} (i) and (ii), respectively.

~\

{\bf Proof of Lemma \ref{lemm:mix0} (i): }

From (2.1)-(2.2) in \cite{2013dependent} and Theorem 1 in \cite{Fang2021HeavyTailedDF}, we have
\begin{align}
& b_{+}(K) := \lim_{t \to \infty}\frac{{\rm pr}(KT_{\phi}>t)}{{\rm pr}(|W_\phi|>t)} = \frac{K}{2},  \\
& b_{-}(K) := \lim_{t \to \infty}\frac{{\rm pr}(KT_{\phi}\le -t)}{{\rm pr}(|W_\phi|>t)} = -\frac{K}{2},
\end{align}
and the limits
$$c_+ := \lim_{K\to \infty} (b_+(K) - b_+(K-1)) = \frac{1}{2}\ \textrm{and}\  c_- := \lim_{K\to \infty} (b_-(K) - b_-(K-1)) = \frac{1}{2}$$
exist. Then the conditions (TB) and (AC) are satisfied. 
Note that $S_d$ is symmetric, condition (CT) in \cite{2011dependent} is satisfied.

From Theorem 3.1 in \cite{2011dependent}, we have 
\begin{align}
\frac{KT_{\phi}}{c_{K,\phi}} \to S\left(1, 0, \frac{\pi}{2}, 0\right)\qquad \textrm{as}\ K\to \infty,
\end{align}
Note that $c_{K, \phi}/K \to 2c_\phi$ as $K\to \infty$, we have
\begin{align}
T_{\phi} \to S\left(1, 0, c_\phi \pi, 0\right)\qquad \textrm{as}\ K\to \infty,
\end{align}
where the characteristic of S(1, 0, $c_\phi \pi$, 0) is $\chi(x) = \exp\{-c_\phi\pi|x|\}$. 

~\

{\bf Proof of Lemma \ref{lemm:mix0} (ii):}

From (2.1)-(2.2) in \cite{2013dependent} and Theorem 1 in \cite{Fang2021HeavyTailedDF}, we have
\begin{align}
& b_{+}(K) := \lim_{t \to \infty}\frac{{\rm pr}(KT_{\phi}>t)}{{\rm pr}(|W_\phi|>t)} = K,  \\
& b_{-}(K) := \lim_{t \to \infty}\frac{{\rm pr}(KT_{\phi}\le -t)}{{\rm pr}(|W_\phi|>t)} = 0. 
\end{align}
And we have the limits
$$c_+ := \lim_{K\to \infty} (b_+(K) - b_+(K-1)) = 1\ \textrm{and}\  c_- := \lim_{K\to \infty} (b_-(K) - b_-(K-1)) = 0$$
exist. Then the condition (TB) is satisfied. 

Next, we consider condition (CT). Actually, (CT) is a centering condition and can be replaced by a location parameter. Specifically, 
from the proof of Theorem 3.1 in \cite{2011dependent}, we have 
\begin{align}
\frac{KT_{\phi}}{c_{K,\phi}} - E\left(\sin\left(KT_{\phi}/c_K\right)\right) \to S\left(1, 1, \frac{\pi}{2}, 0\right)\qquad \textrm{as}\ K\to \infty,
\end{align}
where the characteristic of S(1, 1, $\pi$/2, 0) is $\chi(x) = \exp\{-|x|[0.5\pi+i\textrm{sign(x)}\log|x|]\}$ and $c_{K,\phi}/K \to c_\phi$.
Note that
\begin{align*}
 \mathbb{E}\left(\sin\left(KT_{\phi}/c_{K,\phi}\right)\right) 
&=  \int_{0}^{\infty}\sin(t/c_{K,\phi}) dF(t)  \\
&\sim \int_{0}^{\infty}\sin(x/c_{K,\phi}) dKF_\phi(x) \\
&\sim K \int_{0}^{\infty}\sin(x/c_{K,\phi}) dF_\phi(x) \\
&\sim  K\mathbb{E}\left(\sin\left(W_\phi/K\right)\right).
\end{align*}
Then we have
\begin{align}
T_{\phi} -  \Delta_\phi \to S\left(1, 1, \frac{c_\phi \pi}{2}, 0\right)\qquad \textrm{as}\ K\to \infty,
\end{align}
where $\Delta_\phi = K\mathbb{E}\left(\sin\left(W_\phi/K\right)\right)$.

\end{proof}

{\bf Proofs of Theorem \ref{th:weak} and Corollary \ref{coll:hmp}:}

\begin{proof}

For $T_{\textrm{CCT}}$, we consider $W_{\mathcal{C}}$ following the standard Cauchy distribution, which is symmetric stable. Note that the right tail of $W_{\mathcal{C}}$ is $F_{\mathcal{C}}(x) = 1/2 - \arctan(x)/\pi \sim 1/(\pi x)$, it implies $c_{\mathcal{C}} = 1/\pi$. 
For $T_{\textrm{PCCT}}$, we consider the right tail of $W_{\mathcal{P}}$ is $F_{\mathcal{C}}(x) = 1 - 2\arctan(x)/\pi \sim 2/(\pi x)$, which implies $c_{\mathcal{P}} = 1/\pi$. 
Similarly,  the right tail of $W_{\mathcal{H}}$ is $F_{\mathcal{H}}(x) = 1/x$ for $T_{\textrm{HMP}}$.
From Lemma \ref{lemm:mix0} (ii), we have Theorem \ref{th:weak} and Corollary \ref{coll:hmp}. 

\end{proof}

We note that the asymptotic distribution of $T_{\rm CCT}$ and $T_{\rm HMP}$ are the same as those under the independent assumption. As the following theorem implies, the asymptotic distribution of $T_{\rm PCCT}$ is also the same as the independent case.
\begin{theorem}
\label{theo:bk}
Suppose $p_1,\cdots,p_K$ are independent standard uniform random variables. 
For $\alpha \in (0,1)$, we have
\begin{align}
T_{\rm PCCT}-\Delta_{\mathcal{P}} \to S\left(1, 1, 1, 0\right)\qquad {\rm as}\ K\to \infty,
\end{align}
where the characteristic of $S(1, 1, 1, 0)$ is $\chi(t) = \exp\{-|t|-2i{\rm sign}(t)\ln|t|/\pi\}$, and
\begin{align}
\Delta_{\mathcal{P}} = K\mathbb{E}(W_{\mathcal{P}}/K) = K\int_{0}^{\infty}\sin(x/K)\frac{2}{\pi (x^2+1)}dx, 
\end{align}
where the distribution function of $W_{\mathcal{P}}$ is $F_{\mathcal{P}}(x) = 2\arctan(x)/\pi$ . 
\end{theorem}

\begin{proof}

For $i = 1, \cdots, K$, we note that $\phi(U_i)\sim W_{\mathcal{P}}$ are regularly varying random variables  with the right distribution tail $1-F_{\mathcal{P}}(x) = 1-2\arctan(x)/\pi \sim 2/(\pi x)$. From Theorem 1.8.1 in \cite{1994Stable}, the domain of attraction of stable distributions can be shown by
\begin{align}
\frac{\phi(U_1)+\ldots+\phi(U_K) - c_2}{c_1} \to S(1,1,1,0)
\quad \rm{as }\ K\to \infty,
\end{align}
where the characteristic of $S(1, 1, 1, 0)$ is $\chi(t) = \exp\{-|t|-2i{\rm sign}(t)\ln|t|/\pi\}$ and 
\begin{align}
c_1 = \tan(\pi/2-1/K)  \quad  {\rm and} \quad
c_2 = c_1K\int_{0}^{\infty}\sin(x/c_1)\frac{2}{\pi (x^2+1)}dx.
\end{align}
Note that $c_1/K \to 1$ as $K \to \infty$, we have
\begin{align}
T_{\rm PCCT} -  \Delta_{\mathcal{P}} \to S\left(1, 1, 1, 0\right)\qquad {\rm as}\ K\to \infty,
\end{align}
where 
\begin{align}
\Delta_{\mathcal{P}} = K\mathbb{E}(W_{\mathcal{P}}/K) = K\int_{0}^{\infty}\sin(x/K)\frac{2}{\pi (x^2+1)}dx.
\end{align} 

\end{proof}

\subsection{Proof of Proposition \ref{prop:VAD}}
\begin{proof}
We need to calculate the analytical value of VAD threshold $a_{\mathcal{P}}$, which means 
\begin{align*}
\sup \left\{{\rm pr} \left( M_{\rm{ptan}}(U_{1}, \ldots ,U_{K})\leq a_{\mathcal{P}} \right) \mid U_{1}, \ldots ,U_{K}\in \mathcal{U}\right\} = \alpha
\end{align*}
holds for any $\alpha>0$.


For $U_i \in \mathcal{U},\ i=1,\ldots,K$, we notice that
\begin{align*}
q_{\alpha}\left(M_{\rm{ptan}}(U_1,\ldots,U_K)\right) 
&=  \sup \left\{x : {\rm pr}\left(\psi\left(\frac{1}{K}\sum_{i=1}^K\phi(U_i)\right)\leq x\right) < \alpha\right\} \\
&=  \sup \left\{x : {\rm pr}\left(\frac{1}{K}\sum_{i=1}^K\phi(U_i)\geqslant \phi(x)\right) < \alpha\right\} \\
&=  \sup \left\{x : {\rm pr}\left(\frac{1}{K}\sum_{i=1}^K\phi(U_i) < \phi(x)\right) \geqslant 1-\alpha\right\} \\
&=  \psi\left( \inf \left\{\phi(x) : {\rm pr}\left(\frac{1}{K}\sum_{i=1}^K\phi(U_i) < \phi(x)\right) \geqslant 1-\alpha\right\} \right) \\
&=  \psi\left( \sup\left\{x' : {\rm pr}\left(\frac{1}{K}\sum_{i=1}^K\phi(U_i) < x'\right) < 1-\alpha\right\} \right) \\
&= \psi\left(\frac{1}{K} q_{1-\alpha} \left(\sum_{i=1}^K \phi(U_i)\right) \right).
\end{align*}
The last equality is given by the fact that $\phi(u)$ is a continuous function and strictly monotonic.
From the definition of $a_{\mathcal{P}}(\alpha)$, we have
\begin{align}
a_{\mathcal{P}}(\alpha)
& = \inf \{q_\alpha (M_{{\rm ptan},K}(U_1,\ldots,U_K)\mid U_1,\ldots,U_K\in \mathcal{U})\} \\
& = \psi \left(\frac{1}{K}\sup \left\{q_{1-\alpha} \left(\sum_{i=1}^K \phi(U_i)\right)\mid U_1,\ldots,U_K\in \mathcal{U} \right\}\right).
\end{align}
In our proposed method, $\phi(U_i) \sim F_{\mathcal{P}}$ with $F_{\mathcal{P}}(x) = 2\arctan(x)/\pi$, which has decreasing density on $[0,\infty]$ and the inverse of  $F_{\mathcal{P}}$ is $F_{\mathcal{P}}^{-1}(x) = \tan(\pi x/2)$.

Noticing that $\sup \left\{q_{1-\alpha} \left(\sum_{i=1}^K\phi(U_i)\right)\right\}$ can be regarded as Value-at-Risk in \cite{2013wangVaR} at level $1-\alpha$, the proof of (i) can naturally be derived from the results of robust risk aggregation with their Corollary 3.7.
That is, 
$a_{\mathcal{P}}(\alpha) = \psi_{\rm{ptan}}(H_\alpha(x_K)/K)$,
where $H_\alpha(x) = (K-1)F_{\mathcal{P}}^{-1}(1-\alpha+(K-1)x) + F_{\mathcal{P}}^{-1}(1-x)$ with $x\in(0,\alpha/K)$, and $x_K$ is the unique solution to the equation
$$K\int_x^{\alpha/K}H_\alpha(t){\rm d}t = (\alpha-Kx)H_\alpha(x).$$

Now we prove the approximation of $a_{\mathcal{P}}(\alpha)$ as $\alpha/\{\log(K)a_{\mathcal{P}}(\alpha)\} \to 1$. We notice that
\begin{align*}
 H_\alpha(x) 
=& (K-1)\tan\left\{\frac{\pi}{2}\left[1-\alpha+(K-1)x\right]\right\} +  \tan\left\{\frac{\pi}{2}(1-x)\right\}, \\
\int_x^{\frac{\alpha}{K}}H_\alpha(t)\textrm{d}t 
=& \left\{-\frac{2}{\pi}\ln\cos\left\{\frac{\pi}{2}\left[1-\alpha+(K-1)t\right]\right\} + \frac{2}{\pi}\ln\cos\left\{\frac{\pi}{2}(1-t)\right\}\right\} \Bigg|_{x}^{\frac{\alpha}{K}} \\
=& \frac{2}{\pi}\ln\sin\left\{\frac{\pi}{2}\left[\alpha-(K-1)x\right]\right\} - \frac{2}{\pi}\ln\sin\left(\frac{\pi}{2}x\right).
\end{align*}
Firstly, we suppose $x_K = o(1/K)$. As $K \to \infty$,
\begin{align*}
& \int_{x_K}^{\frac{\alpha}{K}}H_\alpha(t)\textrm{d}t 
\sim  \lim_{K\to +\infty} - \frac{2}{\pi} \ln x_K, \\
& \left(\frac{\alpha}{K} - x_K\right) H_{\alpha}(x_K)
\sim \lim_{K\to +\infty}\frac{\alpha}{K} \frac{2}{\pi x_K}.
\end{align*}
Then we have $x_K \sim \alpha/(K\log K)$ satisfying $x_K = o(1/K)$. So that $x_K$ is the unique solution as
$$x_K = \min\left\{x\in \left(0,\frac{\alpha}{K}\right): K\int_x^{\alpha/K}H_\alpha(t){\rm d}t \le (\alpha-Kx)H_\alpha(x).\right\}$$

Notice that
\begin{align*}
&\lim_{K\to +\infty} H_\alpha(x_K) 
  = \lim_{K\to +\infty} (K-1)\tan\left\{\frac{\pi}{2}\left[1-\alpha+(K-1)x_K\right]\right\} +  \tan\left\{\frac{\pi}{2}(1-x_K)\right\}  
\sim \frac{2}{\pi}\frac{K\log K}{\alpha}, \\
&\lim_{K\to +\infty} a_{\mathcal{P}}(\alpha) 
 =  \lim_{K\to +\infty} \psi_{\rm{ptan}}(H_\alpha\left(x_K\right) /K)
= \lim_{K\to +\infty} \left\{1-\frac{2}{\pi}\arctan\left[H_\alpha(x_K)/K\right]\right\}
 \sim \frac{\alpha}{\log(K)}.
\end{align*}
We prove that $\alpha/\{\log(K)a_{\mathcal{P}}(\alpha)\} \to 1$.

\end{proof}

\subsection{Proof of Theorem \ref{theo:r=-1}}

\begin{proof}
For any $0<\delta_1< 1/K$, there exists $0<\epsilon<1$ such that for all $x\in (0,\epsilon)$
\begin{align}\label{eq:taylor1}
\frac{2-\delta_1}{\pi x}<\tan\left(\frac{\pi}{2}-\frac{\pi}{2}x\right)<\frac{2}{\pi x}.
\end{align}
Let $0< \epsilon' \le \delta_1/2$, then we have
\begin{align}
\inf_{x\in [\epsilon,1]}\left\{\tan\left(\frac{\pi}{2}-\frac{\pi}{2}x\right) - \frac{2}{\pi x} \right\} \geq -\frac{2}{\pi}
\geq -\frac{\delta_1}{\pi \epsilon'}.
\end{align}
Take $(p_1,\ldots,p_K)$ such that $p_{(1)} < \epsilon'$. Let $l = \max\{i=1,\ldots,K:p_{(i)} < \epsilon\}$. And we have
\begin{align*}
\sum_{i=1}^K\tan\left(\frac{\pi}{2}-\frac{\pi}{2}p_{(i)}\right) 
&\geqslant \sum_{i=1}^l\frac{2-\delta_1}{\pi p_{(i)}}
+ \sum_{i=l+1}^K\frac{2}{\pi p_{(i)}} - \frac{(K-l)\delta_1}{\pi p_{(1)}}\\
&\geqslant \sum_{i=1}^K\frac{2-K\delta_1}{\pi p_{(i)}} \\
&= \sum_{i=1}^K\frac{2-K\delta_1}{\pi p_{i}}.
\end{align*}
Similarly, we can show
\begin{align}
\sum_{i=1}^K\tan\left(\frac{\pi}{2}-\frac{\pi}{2}p_{i}\right) 
\leq \sum_{i=1}^K\frac{2}{\pi p_{i}}.
\end{align}
Note that $\psi(y) = 1-2 \arctan(y)/\pi \sim 2/(\pi y)$ as $y\to \infty$. 
For any $\delta_2 \in (0, 1/K)$, there exists $m<0$ such that for all $y\in (-\infty, m)$,
\begin{align}
\frac{2-\delta_2}{\pi y} \leq \psi(y) \leq \frac{2+\delta_2}{\pi y}.
\end{align}
And then for any $(p_1,\ldots,p_K)$ satisfying $p_{(1)} < \epsilon'$,
\begin{align}
\frac{2-\delta_2}{2}M_{-1}(p_1,\ldots,p_K)
\leq M_{\rm{ptan}}(p_1,\ldots,p_K)
\leq \frac{2+\delta_2}{2-K\delta_1}M_{-1}(p_1,\ldots,p_K).
\end{align}
Let $\delta_1,\delta_2\to 0$, we complete $M_{\rm{ptan}}\sim M_{-1}$. From Theorem 2 in \cite{Chen2020TradeoffBV}, we have $ M_{-1}\sim M_{\tan}$, which completes the proof.

\end{proof}

\subsection{Proof of Theorem \ref{theo:three}}

\begin{proof} We proof  Theorem \ref{theo:three} (i) and (ii), respectively.

(i)
We have showed $\alpha/\{\log(K)a_{\mathcal{P}}(\alpha)\} \to 1$ in the proof of Proposition \ref{prop:VAD}. 
From Proposition 4 in \cite{Chen2020TradeoffBV}, for $\alpha\in (0,1/2)$, we have $\mathcal{C}(x) = 1/2 + \arctan(x)/\pi$ and
\begin{align*}
a_{\mathcal{C}}(\alpha) = \mathcal{C}(-H_{\alpha}(x_K)/K),
\end{align*}
where $H_\alpha(x) = (K-1)\mathcal{C}^{-1}(1-\alpha+(K-1)x) +  \mathcal{C}^{-1}(1-x), x\in (0,\alpha/K) $ and $x_K$ is the unique solution to the equation 
\begin{align*}
K\int_{x}^{\alpha/K}H_{\alpha}(t){\rm{d}}t =(\alpha-Kx)H_{\alpha}(x).
\end{align*}

Now we demonstrate the approximation of $a_{\mathcal{C}}(\alpha)$ by $\alpha/\{\log(K)a_{\mathcal{C}}(\alpha)\} \to 1$ as $K\to+\infty$. Notice that
\begin{align*}
 H_\alpha(x) =& (K-1)\mathcal{C}^{-1}(1-\alpha+(K-1)x) +  \mathcal{C}^{-1}(1-x) \\
=& -(K-1)\tan\left\{\frac{\pi}{2}-\pi\left[\alpha-(K-1)x\right]\right\} -  \tan\left\{\frac{\pi}{2}-\pi x\right\}, \\
\int_x^{\frac{\alpha}{K}}H_\alpha(t)\textrm{d}t 
=&\left\{\frac{1}{\pi}\ln\cos\left\{\frac{\pi}{2}-\pi\left[\alpha-(K-1)t\right]\right\} - \frac{1}{\pi}\ln\cos\left(\frac{\pi}{2}-\pi t\right)\right\} \Bigg|_{x}^{\frac{\alpha}{K}} \\
=& -\frac{1}{\pi}\ln\sin\left\{\pi\left[\alpha-(K-1)x\right]\right\} + \frac{1}{\pi}\ln\sin\left(\pi x\right).
\end{align*}

Firstly, we suppose $x_K = o(1/K)$. As $K \to \infty$,
\begin{align*}
& \int_{x_K}^{\frac{\alpha}{K}}H_\alpha(t)\textrm{d}t 
\sim  \lim_{K\to +\infty}  \frac{1}{\pi}\ln x_K, \\
& \left(\frac{\alpha}{K} - x_K\right) H_{\alpha}(x_K)
\sim \lim_{K\to +\infty}- \frac{\alpha}{K} \frac{1}{\pi x_K}.
\end{align*}
Then we have $x_K \sim \frac{\alpha}{K\log K}$ satisfying $x_K = o(1/K)$. So that $x_K$ is the unique solution as
$$x_K = \min\left\{x\in \left[0,\frac{\alpha}{K}\right]: K\int_x^{\alpha/K}H_\alpha(t){\rm d}t \le (\alpha-Kx)H_\alpha(x).\right\}$$

We notice that
\begin{align*}
\lim_{K\to +\infty} H_\alpha(x_K) 
 & = \lim_{K\to +\infty} -(K-1)\tan\left\{\frac{\pi}{2}-\pi\left[1-\alpha+(K-1)x_K\right]\right\} -  \tan\left\{\frac{\pi}{2}-\pi x_K\right\} \\ 
& \sim -\frac{K\log K}{\pi\alpha}, \\
\lim_{K\to +\infty} a_{\mathcal{C}}(\alpha) 
& =  \lim_{K\to +\infty} \mathcal{C}(-H_\alpha\left(x_K\right)/K )
= \lim_{K\to +\infty} K\left\{\frac{1}{2}-\frac{1}{\pi}\arctan\left[H_\alpha(x_K)\right]\right\}
 \sim \frac{\alpha}{\log(K)}.
\end{align*}
We prove that $\alpha/\{\log(K)a_{\mathcal{C}}(\alpha)\} \to 1$, and then $a_{\mathcal{P}} \sim a_{\mathcal{H}} \sim a_{\mathcal{C}} \sim \alpha/\log(K)$.

(ii)
Since $b_{\mathcal{H}}\sim b_{\mathcal{C}}$ has been proven in \cite{Chen2020TradeoffBV}, it is sufficient to show $b_{\mathcal{P}}\sim b_{\mathcal{C}}$ as $\alpha\to0^+$, where $b_{\mathcal{P}}$ and $b_{\mathcal{C}}$ are the VWD thresholds for our proposed method and the Cauchy combination test, respectively.

From the generalized central limit theorem, 
\begin{align}
b_{\mathcal{H}}(\alpha)\sim \left(\frac{c_1^{\mathcal{H}}p_S(1-\alpha)+c_2^{\mathcal{H}}}{K}\right)^{-1}
\quad {\rm as}\ K\to \infty,
\end{align}
where $c_1^{\mathcal{H}} = \pi K/2$ and $c_2^{\mathcal{H}} = c_1^{\mathcal{H}}K\int_{1}^{\infty}\sin(x/c_1^{\mathcal{H}})x^{-2}dx$. We note that $\Delta_{\mathcal{P}} = o(K)$, $c_2^{\mathcal{H}}/K = o(K)$ and $p_S(1-\alpha) \to \infty$ as $\alpha \to 0^+$. Then we have 
\begin{align}
\lim_{\alpha\to 0^+, K\to \infty}b_{\mathcal{P}}(\alpha)
= \lim_{\alpha\to 0^+, K\to \infty}
\left(\frac{\pi}{2}p_S(1-\alpha)\right)^{-1}
= \lim_{\alpha\to 0^+, K\to \infty}b_{\mathcal{H}}(\alpha),
\end{align}
\cite{2015stablecompute} showed an approximation to quantiles of the stable distributions $S(\alpha, 1, 0, 1)$ as
\begin{align}
1 - F_\alpha(x) \to \frac{1}{\pi\alpha}2\Gamma(\alpha+1)\sin\left(\frac{\pi}{2}\alpha\right)y^{-\alpha} \qquad {\rm as}\ x\to \infty,
\end{align}
where $F_\alpha(x)$ is the distribution function of $S(\alpha, 1, 0, 1)$ and $x = y + 2\log(y)/\pi$ as $\alpha \to 1$. Let $1 - F_\alpha(x_\alpha) = \alpha$ and then $x_\alpha = p_S(1-\alpha)$, we note that
\begin{align}
p_S(1-\alpha) \to \frac{2}{\pi\alpha} \quad{\rm as}\ \alpha\to 0^+.
\end{align}
From Theorem \ref{th:weak}, we notice $T_{\rm CCT}  \stackrel{d}{\to} W_{\mathcal{C}}$ as $K\to \infty$ so that $b_{\mathcal{C}}(\alpha) \to \alpha$. 
Then we have $\lim_{\alpha\to 0^+, K\to \infty}b_{\mathcal{P}}(\alpha) = \alpha = b_{\mathcal{C}}(\alpha) = \alpha$ and complete the proof.

\end{proof}

\subsection{Proof of Theorem \ref{theo:power}}

\begin{proof} We proof Theorem \ref{theo:power} (i) and (ii), respectively.

(i) To derive ${\rm pr}(T_{\rm{PCCT}}>t_{\alpha/2})\ge {\rm pr}(T_{\rm{CCT}}>t_{\alpha})$, that is 
\begin{align}
{\rm pr}\left\{\sum_{i=1}^{K} \frac{1}{K} \tan \left(\frac{\pi}{2}-\frac{\pi}{2}p_i\right)  > \tan \left(\frac{\pi}{2}-\frac{\pi}{2}\alpha\right) \right\} \ge {\rm pr}\left\{\sum_{i=1}^{K} \frac{1}{K} \tan \left(\frac{\pi}{2}-\pi p_i\right)  > \tan \left(\frac{\pi}{2}-\pi\alpha\right) \right\},
\end{align}
we can prove that  $\alpha_{P,K} \le \alpha_{C,K}$ holds for any $\{p_1,p_2, \cdots, p_K\}$ and any $K\in\mathbb{N}$, where $\alpha_{P,K}$ and $\alpha_{C,K}$ satisfying 
\begin{align}
&\sum_{i=1}^{K} \frac{1}{K} \tan \left(\frac{\pi}{2}-\frac{\pi}{2} p_i\right)  = \tan \left(\frac{\pi}{2}-\frac{\pi}{2}\alpha_{P,K}\right), \\
&\sum_{i=1}^{K} \frac{1}{K} \tan \left(\frac{\pi}{2}-\pi p_i\right)  = \tan \left(\frac{\pi}{2}-\pi\alpha_{C,K}\right).
\end{align}
We consider $F_1(x) = \tan \left(\pi/2-\pi x/2\right)$ and $F_2(x) = \tan \left(\pi/2-\pi x\right)$, $x\in\left(0,1\right)$.
Then we have $T_{\rm{PCCT}} = \sum_{i=1}^K F_1(p_i)/K$ and $T_{\rm{CCT}} = \sum_{i=1}^K F_2(p_i)/K$.

We now finish the proof by mathematical induction for $K\in\mathbb{N}$, which can be divided into two steps. In the first step, we are to prove a more general result for $K = 2$.  For any $p_1<p_2$ and $\omega = 1/2$, we have 
$F_1(\alpha_{P,K}) = \omega F_1(p_1) + (1-\omega)F_1(p_2)$ and 
$F_2(\alpha_{C,K}) = \omega F_2(p_1) + (1-\omega)F_2(p_2)$.
Then
\begin{align}
\frac{1}{\omega} - 1 = \frac{F_1(p_1) - F_1(\alpha_{P,K})}{F_1(\alpha_{P,K}) - F_1(p_2)}  = \frac{F_2(p_1) - F_2(\alpha_{C,K})}{F_2(\alpha_{C,K}) - F_2(p_2)}.
\end{align}
Notice that $F_1'(x)  < 0$ and $F_1''(x) > 0$, we try to prove
\begin{align}
\frac{F_1(p_1) - F_1(\alpha_{C,K})}{F_1(\alpha_{C,K}) - F_1(p_2)} \ge \frac{F_2(p_1) - F_2(\alpha_{C,K})}{F_2(\alpha_{C,K}) - F_2(p_2)},
\end{align}
and then $\alpha_{P,K} \le \alpha_{C,K}$. It is equivalent to prove 
\begin{align}\label{eq:f1f2}
\left[\frac{F'_1(x)}{F'_2(x)}\right]^{'} \le 0.
\end{align}
Notice that for $x\in(0,1)$
\begin{align*}
&F'_1(x) = -\frac{\pi}{2}\sec^2\left(\frac{\pi}{2} - \frac{\pi}{2}x\right) < 0, \\
&F'_2(x) = -\pi \sec^2\left(\frac{\pi}{2} - \pi x\right) < 0, \\
&F''_1(x) = \frac{\pi^2}{2}\sec^2\left(\frac{\pi}{2} - \frac{\pi}{2}x\right)\tan\left(\frac{\pi}{2} - \frac{\pi}{2}x\right), \\
&F''_2(x) = 2\pi^2 \sec^2\left(\frac{\pi}{2} - \frac{\pi}{2}x\right)\tan\left(\frac{\pi}{2} - \pi x\right) ,
\end{align*}
and then we have
\begin{align}
\frac{F''_1(x)}{F'_1(x)} = -\pi \tan\left(\frac{\pi}{2} - \frac{\pi}{2}x\right) 
< - 2\pi \tan\left(\frac{\pi}{2} - \pi x\right) 
= \frac{F''_2(x)}{F'_2(x)}.
\end{align}
Therefore, $F''_1(x) F'_2(x) - F'_1(x) F''_2(x) < 0$ and then \ref{eq:f1f2} holds. Now $\alpha_{P,K} \le \alpha_{C,K}$ holds for any $0<\omega<1$ and $K=2$.

In the second step, suppose $\alpha_{P,K-1} \le \alpha_{C,K-1}$ hold, where 
$F_1(\alpha_{P,K-1}) = \sum_{i=1}^{K-1}F_1(p_i)/(K-1)$
and $F_2(\alpha_{C,K-1}) = \sum_{i=1}^{K-1}F_2(p_i)/(K-1)$. 
We aim to verify $\alpha_{P,K} \le \alpha_{C,K}$, where $F_1(\alpha_{P,K}) = (K-1) F_1(\alpha_{P,K-1})/K + F_1(p_K)/K$
and $F_2(\alpha_{C,K}) = (K-1) F_2(\alpha_{C,K-1})/K + F_2(p_K)/K$.
Let $\alpha'$ satisfy
\begin{align}
F_1(\alpha') = \frac{K-1}{K} F_1(\alpha_{C,K-1}) + \frac{1}{K}F_1(p_K).
\end{align}
Recalling the general results for $K=2$, we have $\alpha' \le \alpha_{C,K}$, and $\alpha_{P,K-1} \le \alpha_{C,K-1}$ leads to $F_1(\alpha_{P,K}) \geq F_1(\alpha')$.
Then $\alpha_{P,K} \le \alpha' \le \alpha_{C,K}$.

In summary, $\alpha_{P,K} \le \alpha_{C,K}$ holds for any $K\in\mathbb{N}$, which completes the proof.



(ii) If the Bonferroni method works, we have 
\begin{align}
Kp_{(1)}< \alpha = \lim_{K\to \infty,\alpha\to 0^+}b_{\mathcal{H}}(\alpha).
\end{align}
Note that
\begin{align}
M_{-1}(p_1,\ldots,p_K) = \left(\frac{1}{K}\sum_{i=1}^K \frac{1}{p_i}\right)^{-1} 
\leq \left(\frac{1}{K}\frac{1}{p_{(1)}}\right)^{-1} = Kp_{(1)}.
\end{align}
As $K\to \infty$ and $\alpha\to 0^+$, we have
\begin{align}
\lim_{K\to \infty,\alpha\to 0^+}M_{\rm{ptan}}(p_1,\ldots,p_K) 
= \lim_{K\to \infty,\alpha\to 0^+}M_{-1}(p_1,\ldots,p_K) 
\leq Kp_{(1)}.
\end{align}
And we have $\lim_{K\to \infty,\alpha\to 0^+}M_{\rm{ptan}}<\lim_{K\to \infty,\alpha\to 0^+}b_{\mathcal{H}}(\alpha)$ falling into the critical region, which completes the proof.
\end{proof}

\end{document}